\documentclass[namedreferences]{solarphysics}

\usepackage[hyperref,optionalrh,natbib]{spr-sola-addons} % For Solar Physics 
\usepackage{graphicx}        % For eps figures, newer & more powerfull
\usepackage{color}           % For color text: \color command
\usepackage{breakurl}        % For breaking URLs easily trough lines

\DeclareRobustCommand{\ion}[2]{%
\relax\ifmmode
\ifx\testbx\f@series
{\mathbf{#1\,\mathsc{#2}}}\else
{\mathrm{#1\,\mathsc{#2}}}\fi
\else\textup{#1\,{\mdseries\textsc{#2}}}%
\fi}
\newcommand{\farcs}{\hbox{$.\!\!^{\prime\prime}$}}
\usepackage{natbib} 
\bibpunct{(}{)}{;}{a}{}{,}

\begin{document}

\begin{article}

\begin{opening}

\title{Spectroscopy at the solar limb: II. Are spicules heated to coronal temperatures ?}

   \author{C.~\surname{Beck}$^{1}$ \sep R.~\surname{Rezaei}$^2$ \sep K.~G.~\surname{Puschmann}$^{3}$\sep D.~\surname{Fabbian}$^{4,5,6}$}
        
   \runningtitle{Spectroscopy at the solar limb: II. Are spicules heated to coronal temperatures ?}
  \runningauthor{C. Beck, R. Rezaei, K.~G. Puschmann, D. Fabbian}

   \institute{$^1$: National Solar Observatory (NSO), 
     %3010 Coronal Loop, 
     88349 Sunspot, New Mexico, USA 
     \url{cbeck@nso.edu}\\
  $^2$:  Kiepenheuer-Institut f\"ur Sonnenphysik (KIS),
     %Sch\"oneckstr. 6, 
     79104 Freiburg, Germany\\
     %\url{rrezaei@kis.uni-freiburg.de}\\
     $^3$:  %ESA/ESOC Darmstadt, 
     %Robert-Bosch-Str. 5, 
     Martinstr.~64, 64285 Darmstadt, Germany\\ %64293 Darmstadt, Germany\\
%     \url{Klaus.Gerhard.Puschmann@esa.int}\\
    $^4$: Instituto de Astrof\'{\i}sica de Canarias
     (IAC), 
     %V{\'i}a Lact{\'e}a,
     38205 La Laguna, Tenerife, Spain\\
%       \url{damian@iac.es}\\
      $^5$: Departamento de Astrof{\'i}sica, Universidad de La Laguna (ULL), 38206
     La Laguna, Tenerife, Spain\\
    $^6$:   Max-Planck-Institut f\"ur
        Sonnensytemforschung (MPS), 
  37077 G\"ottingen, Germany}
 %, Justus-von-Liebig-Weg 3
\begin{abstract}

Spicules of the so-called type II were suggested to be relevant for coronal
heating because of their ubiquity on the solar surface and their eventual
extension into the corona. We investigate whether solar spicules are heated to
transition-region or coronal temperatures and reach coronal heights ($\gg
6$\,Mm) using multi-wavelength observations of limb spicules in different
chromospheric spectral lines (\ion{Ca}{ii} H, H$\epsilon$, H$\alpha$,
\ion{Ca}{ii} IR at 854.2\,nm, \ion{He}{i} at 1083\,nm) taken with
slit-spectrographs and imaging spectrometers. We determine the line width of
spectrally resolved line profiles in individual spicules and throughout the
field of view and estimate the maximal height that different types of off-limb
features reach. We derive estimates of the kinetic temperature and the
non-thermal velocity from the line width of spectral lines from different chemical elements. We find that most regular -- i.e., thin and elongated --
spicules reach a maximal height of about 6\,Mm above the solar limb. The
majority of features found at larger heights are irregularly shaped with a
significantly larger lateral extension -- of up to a few Mm -- than
spicules. Both individual and average line profiles in all spectral lines show
a decrease in their line width with height above the limb with very few
exceptions. Both the kinetic temperature and the non-thermal velocity decrease
with height above the limb. We find no indications that the spicules in
  our data reach coronal heights or transition-region or coronal
temperatures. 
\end{abstract}

\end{opening}

%\maketitle he non-thermal velocity decreases from about 30\,km\,s$^{-1}$ near the limb to 5 to 15\,km\,s$^{-1}$ at the uppermost heights reached by spicules. 

\section{Introduction}
Most of the processes on the solar surface, with the exception of the convective energy transport and solar oscillations, are driven by magnetic fields. At the 
photospheric level, the gas density is high enough to make the kinetic
pressure dominant over the magnetic pressure. In contrast, the magnetic energy
density is larger than the kinetic one in the chromosphere and the
corona. Whereas the spatial structuring in the photosphere is given by the
granulation pattern, the shape of the solar chromosphere is markedly
different. One key component of the chromosphere are the so-called spicules:
hair-like, thin, elongated features observed at the solar limb in strong
chromospheric lines such as H$\alpha$ (e.g., \citeauthor{roberts1945}
\citeyear{roberts1945} or the reviews of \citeauthor{beckers68}
\citeyear{beckers68} (BE68), \citeauthor{sterling00} \citeyear{sterling00} and
\citeauthor{tsiropoula+etal2012} \citeyear{tsiropoula+etal2012}).  Some
spicules are bright in extreme ultraviolet (EUV) lines as well as in
H$\alpha$, indicating an extension up to coronal heights. Spicules are
transient features that apparently shoot up from the solar limb to a
  height of a few Mm \citep[e.g.,][]{zirker1962}, to disappear from sight again after some tens of seconds to a few minutes \citep[BE68;][]{depontieu_hansteen_etal07}. Because of possible projection effects, it is not clear if the apparent motions seen at the limb are caused by mass motions or waves. There are controversial observational arguments whether spicules fade away \textit{in situ} or return to the solar surface \citep{suematsu_etal95,pasachoff+etal09,sterling+etal2010,anan+etal2010,pereira+etal2012,zhang+etal2012}. \citet{pasachoff+etal1968} and \citet{depontieu+etal2012} suggested that spicules undergo twisting and torsional motions while they evolve. 

On the solar disc, dark and bright mottles and fibrils are believed to be the counterpart of limb spicules \citep{beckers72}. There is no canonical proof that mottles or fibrils are spicules seen on-disc, but they should at least be related \citep{grossman_schmidt92,tsirop_schmieder,suematsu98,zachariadis_etal99,langangen+etal2008,rouppe_leen_etal09}. Mottles cluster at the boundary of supergranular cells. The most reliable identification of the on-disc counterpart of spicules is through the corresponding line-of-sight (LOS) velocities \citep[e.g.,][]{sekse+etal2012,sekse+etal2013}. There is no agreement if there are any spicules over plage regions \citep[][but see also \citeauthor{anan+etal2010} \citeyear{anan+etal2010}]{shibata_suematsu82,zirin_book, depontieu_hansteen_etal07}. 

The typical length of spicules is from 5--10\,Mm, while their width ranges
from 1\,Mm down to the resolution limit of the respective observations, i.e., as small as 0\farcs1 \citep{beckers72,nishikawa88,suematsu_etal07,suematsu+etal2008}.  Hence, the aspect ratio of the hair-like spicular structures is about 10 or more. They are usually inclined with respect to the local vertical by about 10--40 degree \citep{pasachoff+etal09}. 
In polar regions, spicules are close to vertical and show a larger extension
(BE68). The ``traditional'' spicules have a lifetime of some 5 to 15 min. In
H$\alpha$, they show a velocity of some 25\,km\,s$^{-1}$ and typical
chromospheric temperatures of about 10\,000\,K \citep[][]{zirker1962a,beckers72,matsuno_etal1988,makita2003}. At low spatial resolution, they appear to emanate from uni-polar regions \citep{suematsu_etal95}. Finally, it is not clear whether spicules rise smoothly or intermittently. 

The spectral signature of spicules varies in the red and blue wings of the
H$\alpha$ line. \cite{beckers72} noted that this can be due to a variation in
Doppler shift or line width \citep[see also][]{shoji+etal2010}. The height and
thickness of spicules changes between, e.g., H$\alpha$ and
\ion{Ca}{ii}\,H\,\&\,K. This difference is the joint action of different
spatial resolution and smoothing effects at different wavelengths as well as a
physical difference in the response of the two lines. 

The interest in spicules has revived thanks to new observations with improved
spatial resolution, higher temporal cadence or enhanced spectropolarimetric
sensitivity. One major source for data of the first two categories is the
Solar Optical Telescope \citep{tsunami+etal2008} and its instrumentation
on-board the Hinode satellite \citep{kosugi+etal2007}. Observations with the
0.3-nm-wide \ion{Ca}{ii}\,H interference filter of Hinode have been used to
address the structure and evolution of spicules and larger-scale filaments and
prominences near the limb \citep[e.g.,][]{suematsu_etal07,berger+etal2008},
although the two-dimensional (2D) imaging data provides only apparent motions
in the sky plane in intensity images. The Hinode Ca filter actually also
covers the chromospheric emission of the H$\epsilon$ line that contributes up
to 30\,\% to the filter intensity near the solar limb
\citep{beck+etal2013c}. Another source of fast, high-resolution imaging at
multiple wavelengths is the Rapid Oscillations in the Solar Atmosphere
\citep[ROSA][]{jess+etal2010} instrument at the Dunn Solar Telescope (DST;
Sunspot, USA) which has been used to study, e.g., oscillations inside of spicules \citep{jess+etal2012}.

% (previously: G\"ottingen)
The information content of  high-resolution imaging data was extended to
high-resolution imaging spectroscopy with the Interferometric BIdimensional
Spectrometer \citep[IBIS;][]{cavallini2006,reardon+cavallini2008} at the DST, the GREGOR Fabry-P\'erot Interferometer
  \citep[GFPI;][]{puschmann+etal2006,puschmann+etal2007,puschmann+etal2012b,puschmann+etal2012a,puschmann+etal2012c,puschmann+etal2013} at the German Vacuum Tower Telescope \citep[VTT; Tenerife, Spain;][]{schroeter+soltau+wiehr1985}, or the CRISP instrument \citep{scharmer+etal2008} at the  Swedish 1-m Solar Telescope (SST; La Palma, Spain). Such data of high spatial resolution in H$\alpha$ and the \ion{Ca}{ii} IR line at 854\thinspace nm have been used to determine the properties of spicules or related features off the limb \citep{pasachoff+etal09,depontieu+etal2012} or on the solar disc \citep{langangen+etal2008,sanchezandrade+etal2008,rouppe_leen_etal09,sekse+etal2013}. 

Data with enhanced polarimetric sensitivity were provided by the Advanced
Stokes Polarimeter \citep[ASP;][]{skumanich+etal1997} at the DST in the
\ion{He}{i} D$_3$ line
\citep[e.g.,][]{casini+etal2003,lopezariste+casini2005}, the Tenerife Infrared
Polarimeter \citep[TIP;][]{martinez+etal1999,collados+etal2007} at the VTT in the \ion{He}{i} line at 1083\,nm \citep[e.g.,][]{centeno+etal2010,martinezgonzalez+etal2012}, the SPINOR instrument \citep{socas+etal2006} at the DST in \ion{He}{i} at 1083\,nm and the \ion{Ca}{ii} IR lines \citep{socasnavarro+elmore2005}, the polarimetric mode of the THEMIS telescope \citep{lopezariste+etal2000,paletou+etal2001}, or the Z{\"u}rich Imaging Polarimeter \citep[ZIMPOL;][]{gandorfer+etal2004} at the Gregory-Coud\'e Telescope in Locarno in the \ion{He}{i} D$_3$ line \citep[e.g.,][]{ramelli+etal2006}.

The new data allowed old observational and theoretical results on
spicules to be revised. With the spectropolarimetric observations, the magnetic field strength in spicules could be determined to be between 10\,G and 50\,G
\citep[][]{trujillobueno+etal2005,centeno+etal2010}. \citet{orozco+etal2015}
found a decrease in magnetic field strength from 80\,G at the limb to 30\,G
at a height of 3\,Mm above it. \citet{depontieu_macintosh_etal07} introduced
two types of spicules based on their lifetime \citep[15\,min for type\,I
vs.~2\,min or less for type\,II; see also][]{rouppe_leen_etal09}, where short-lived
spicules are also generally thinner and show faster apparent velocities than
traditional spicules, although already BE68 described two types of spicules
differing in their line widths. \citet{depontieu_macintosh_etal07} stated that
most of the type II spicules do not show a descent, but fade from sight
\textit{in situ}, which could be caused by a rapid heating to transition-region and coronal temperatures. It is not clear at present whether these new
and old two types correspond eventually to the same classification and what
the exact differences between type I and type II spicules
are. \citet{avery1970} explained the two types of spicules defined by BE68 as
being solely caused by different (or absent)
rotation. \citet{pereira+etal2012} suggested that the two types of spicules of
BE68 would fall into the new type I category as defined by
\citet{depontieu_macintosh_etal07}, whereas the new type II spicules should
have been undetectable at the temporal and spatial resolution of older
observations \citep[cf.~also][]{pereira+etal2013}. \citet{zhang+etal2012}
re-analyzed the data used by \citet{depontieu_macintosh_etal07} and found both
an ascending and descending phase for most of their spicule examples. They
questioned the existence of type II spicules as being somehow different from
the classical type I spicules \citep[see also the discussion in][]{skogsrud+etal2015}.
%, but we note that any type of rigid rotation should produce a significant twist of magnetic field lines.

There are suggestions that different mechanisms work on different type of
spicules \citep{hammer_etal08,martinez+etal2009} and various theoretical
models have been proposed \citep[][and references therein]{sterling00}. Such
models have to provide a source of energy to support spicules against gravity,
to accelerate them upwards, and to explain their elongated shape and temporal
evolution. Candidates for the energy source are photospheric impulsive events
\citep{hollweg1982,suematsu_etal1982,depontieu_etal04} and energy sources that are related to
magnetic fields, such as Alfv\'en waves
\citep[e.g.,][]{depon_haer98,kudoh+shibata1999,hansteen_dipon_etal06,rouppe_dipon_etal07},
small-scale reconnection events \citep{heggland+etal2009,yurchyshyn+etal2013}
or localized currents that accelerate material by the Lorentz force
\citep{martinez+etal2011a,goodman2012}. There are indications that the height
extension of spicules is affected by the transition-region height and vice
versa \citep{shibata_suematsu82,guerreiro+etal2013}. Magneto-hydrodynamical
(MHD) wave models of spicules
\citep{kulidzanishvili+zhugzhda1983,kukhianidze+etal2006,he_etal09} are
motivated by observations of oscillations and observed Doppler shifts (the
so-called line tilt) that again date back to BE68 and before. Despite a lot of
observational constraints, no canonical model for spicules that matches their
principal observed dynamical properties could be derived . 

Because spicules, or more precisely, type II spicules according to the new
definition, were found to be abundantly present on the solar surface, they
potentially could transport a significant amount of energy into the corona
\citep{depontieu+etal2009}. As discussed by \citet{judge+etal2012}, this idea
of a coronal energy input from the chromosphere was suggested already earlier
on \citep{athay+holzer1982,athay2000}, but there were only indirect proofs for
this process, e.g., the amount and direction of (vertical) net mass flows
\citep{pneuman+knopp1978}. Most of the argumentation for a relation between
the chromospheric spicules and the transition region or corona was based on
the dynamic evolution and the dynamic properties of spicules \citep[LOS
velocities, acceleration/deceleration on disc centre and near the limb;
e.g.,][]{mcintosh+depontieu2009}. \citet{madjarska+etal2011} were unable to
find coronal counterparts to three large-scale macrospicules
\citep[cf.][]{bohlin+etal1975,pike+harrison1997,kamio+etal2010,scullion+etal2010,murawski+etal2011}
seen in \ion{Ca}{ii} H imaging from Hinode. \citet{pereira+etal2014} and
\citet{rouppevandervoort+etal2015} found counterparts to spicules in
transition-region lines at the limb and on the disc, respectively.

Here, we investigate a possible connection of spicular material to atmospheric
layers above the chromosphere by determining the maximal height above the
solar limb attained by different kinds of features  and by deriving the height dependence of the line width in various chromospheric spectral lines. Assuming a heating process from chromospheric (5000 to 20,000\,K) to transition-region temperatures ($>10^5$\,K) during their rise, the line width in resolved spectra should increase with height if the upper end of spicules gets heated.

Section \ref{sec_obs} describes the various data sets covering
several chromospheric spectral lines (\ion{Ca}{ii} H, H$\epsilon$, H$\alpha$,
\ion{Ca}{ii} IR at 854.2\,nm, \ion{He}{i} at 1083\,nm) in one-dimensional (1D)
and two-dimensional (2D) spectroscopy. The data reduction and the quantities derived from the
spectra are explained in Sect.~\ref{sec_datared_ana}. The results of
Sect.~\ref{sec_results} are summarized in Sect.~\ref{sec_summ} and discussed in Sect.~\ref{sec_disc}. Section \ref{sec_concl} provides our conclusions. Appendix \ref{app_ha} shows a few more examples of time-series of imaging spectroscopy in H$\alpha$. Appendix \ref{signi} discusses the significance limits of the profiles and the analysis approach while Appendix \ref{slit_add} shows additional examples of spectrograph observations.

%Appendix \ref{app_obs} contains a detailed description of the instrumental setups during the observations. The specific methods to obtain a flat-field correction are explained in Appendix \ref{flat_field}, while 
\section{Observations\label{sec_obs}}
For the simultaneous observations of up to five chromospheric spectral lines
in four different wavelength regimes from the near-ultraviolet (UV) to the
near-infrared (IR), we used different combinations of the post-focus instruments
available at the VTT at that time: the main Echelle spectrograph for spectropolarimetric observations of \ion{He}{i} 1083\,nm with TIP and simultaneous spectroscopy of \ion{Ca}{ii} IR at 854.2\,nm with a PCO 4000 camera; the Triple Etalon SOlar Spectrometer \citep[TESOS;][]{kentischer+etal1998,tritschler+etal2002} for imaging spectroscopy of H$\alpha$; the POlarimetric LIttrow Spectrograph \citep[POLIS;][]{beck+etal2005b} for spectroscopic observations of \ion{Ca}{ii} H, H$\epsilon$ and H$\alpha$. The two combinations of the instruments that we used are described in detail in \citet{beck+rezaei2012}. Their main difference is the usage of TESOS for 2D spectroscopy in H$\alpha$  in 2010 (setup 1), while in 2011 all lines were observed with slit-spectrographs (setup 2). Table \ref{obs_specs} lists the spectral and spatial characteristics of the data that only vary for \ion{Ca}{ii} H and H$\alpha$.

%\textbf{\textcolor{red}{The third instrument we used is GFPI. We analysed and will present in this work data for H$\alpha$ obtained in 2005 at the VTT (setup 3).}}

In setup 1, TESOS was run without direct synchronization to the scanning. We set it to a continuous observing mode as soon as the adaptive optics (AO) was locked near the limb. The cadence of TESOS was either about 20 or 30 s, depending on whether a second line (\ion{Mg}{i} at 517\,nm) was recorded in addition to H$\alpha$. The TESOS field of view (FOV) was circular with a diameter of 40$^{\prime\prime}$.
%Each scan was programmed to last about 20 to 30 min to be able to execute different modes on the same solar location and on the same day within the usually one-hour long time window of good seeing conditions on each day.  %Observing the \ion{Mg}{i} was, however, dropped after one day because the permanent motion of the TESOS filter wheel to switch between the two lines created additional flat field problems. 
In both setups, we aimed for observations of spicules in three different observing modes of the spectrograph instruments: large-area maps, time-series (ts), and long-integrated ($>30\,$s) small-area maps. All three types of observations were done with the slit parallel and perpendicular to the limb. The fastest cadence in the slit-spectrograph data was 80 seconds and the integration time per scan step, $t$, varied between 6 seconds and 2 minutes. 

We finally selected only a subset of five specific observations for the
current study. The corresponding settings are listed in Table
\ref{tab_obs}. In all observations, the chromospheric spectral lines of
\ion{He}{i} at 1083\,nm, \ion{Ca}{ii} IR at 854.2\,nm, H$\alpha$ at 656\,nm
and \ion{Ca}{ii} H at 396.85\, nm were observed simultaneously. In setup 1
(observations Nos.~1 to 4), also H$\epsilon$ at 397\,nm was covered in the
POLIS data and the H$\alpha$ line was recorded with a 2D spectrometer, but
only sampled within a limited spectral range. In setup 2 (observation No.~5),
the spectral range around H$\alpha$ was extended because of observing it with
a slit-spectrograph, but H$\epsilon$ was not covered anymore in the
\ion{Ca}{ii} H spectra because of using the default Ca CCD of POLIS.
%(cf.~Table~\ref{obs_specs}) \textbf{\textcolor{red}{([DF: ] To what info in Table 2 are you referring here? I would simply remove this.)}}
\begin{table*}
\caption{Spatial and spectral characteristics of the data.\label{obs_specs}}
\begin{tabular}{ccccc}
line & dispersion & wavelength & slit width & spat. sampl. \cr
     & pm/pixel  & nm         &$^{\prime\prime}$ & $^{\prime\prime}$/pixel \cr\hline\hline
\multicolumn{5}{c}{Setup 1 \& 2: TIP \& Echelle spectrograph}\cr\hline
\ion{He}{i} & 1.1 & 1082.34\,--\,1083.45 & 0.36 & 0.18 \cr
\ion{Ca}{ii} IR & 0.82 &  853.45\,--\,855.10 & 0.36 & 0.18\cr\hline
\multicolumn{5}{c}{Setup 1: TESOS and POLIS}\cr\hline
H$\alpha$$^{1}$ & 4.9 &  656.16\,--\,656.39 & -- & 0.09\cr
\ion{Ca}{ii} H$^{2}$ & 1.45 &  395.86\,--\,398.75 & 0.5 & 0.22\cr\hline
\multicolumn{5}{c}{Setup 2: POLIS}\cr\hline
H$\alpha$$^{2}$ & 2   &  654.22\,--\,657.61 & 0.5 & 0.22 \cr
\ion{Ca}{ii} H$^{3}$ & 1.92 &  396.34\,--\,396.95 & 0.5 & 0.292\cr\hline
\multicolumn{5}{c}{Setup 3: GFPI}\cr\hline
H$\alpha$ & 11.1 &  656.17\,--\,656.40 & -- & 0.11\cr
\end{tabular}\\
$^{1}$: TESOS\,\, $^{2}$: PCO 4000 in POLIS\,\, $^{3}$: default POLIS CCD
\end{table*}
%656.173 ... 656.395
% 11.1 pm  0.11'' / pixel

\begin{table*}
\caption{Settings of the observations.\label{tab_obs}}
\begin{tabular}{c|ccccc}
No. & 1 & 2 & 3 & 4 & 5 \cr\hline\hline
date &  30/06/2010 & 01/07/2010 &  01/07/2010 &  02/07/2010 &12/04/2011\cr
time [UT] &  07:48\,--\,08:08 & 07:56\,--\,08:24& 08:59\,--\,09:19&08:11\,--\,08:24  &07:50\,--\,08:14\cr
pos. $x^{\prime\prime}/y^{\prime\prime}$ & $-864/349$ & $-868/293$ & $-868/293$ & $-354/-836$ & $-933/172$\cr
type & QS/map & QS/map & QS/map & QS/ts & AR/map \cr
$t$ [s] & 10 & 60 & 6 & 6 & 8\cr
steps$^{1}$ &  100$\times$0.3 & 25$\times$0.3  & 150$\times$0.3 & 20$\times$0.5$\times$5 & 150$\times$0.36 \cr
duration [min]& 20 & 28 & 19 & 13 & 25\cr
cadence [s] & 30$^2$ & 20$^2$ & 20$^2$ & 156$^3$ & --\cr
angle$^4$ [deg] & 90 & 90 & 90 & 0 & 90\cr
setup$^5$ & 1 & 1 & 1 & 1 & 2\cr
\cr
No. & 6 &  &  &  &  \cr\hline\hline
date &  04/05/2005 & &  & & \cr
time [UT] &  08:47\,--\,09:02 & \cr
pos. $x^{\prime\prime}/y^{\prime\prime}$  &  south pole \cr
type &  QS/ts & \cr
$t$ [ms]  & 20 \cr
steps  & 39  \cr
duration  [min]& 15 \cr
cadence [s] & $\sim 20$ \cr
setup$^5$ & 3 \cr
\end{tabular}\\
$^{1}$: number of steps\,$\times$\,step width in arc-seconds ($\times$\,number of repetitions of the scan)\,\, $^{2}$: {cadence of TESOS spectra}\,\, $^{3}$: cadence of slit-spectrograph spectra\,\, $^4$: angle relative to the solar limb\,\, $^5$: setup 1: TIP, TESOS and POLIS; setup 2: TIP and POLIS; setup 3: GFPI
\end{table*}

In addition to the multi-line observations described above, we used a
time-series of about 15 min obtained with the GFPI on 04/05/2005 at the VTT
(setup 3) in only the H$\alpha$ line. The spectral line was sampled in 21
steps of 11 pm width with a cadence of about 22\,s
\citep{puschmann+etal2006}. The spectra were reconstructed with the
Multi-Object Multi-Frame Blind Deconvolution Technique
\citep[MOMFBD;][]{loefdahl2002,vannortetal05} as part of the Imaging
Spectropolarimetric Parallel-Organized Reconstruction Data Pipeline (ISPOR-DP),
the GFPI data pipeline described in \citet{puschmann+beck2011}. These data have the highest spatial resolution within our sample. The spectra covered a similar wavelength range as the H$\alpha$ spectra from TESOS but had a coarser spectral sampling.

\begin{figure}
%\resizebox{3.cm}{!}{\includegraphics{300610_mdi.psmagn.ps}}\hspace*{.5cm}\resizebox{3.cm}{!}{\includegraphics{010710_mdi.psmagn.ps}}\hspace*{.5cm}\resizebox{3.cm}{!}{\includegraphics{020710_mdi.psmagn.ps}}\hspace*{.5cm}\resizebox{3.cm}{!}{\includegraphics{120411_mdi.psmagn.ps}}\\$ $\\
%\resizebox{3.cm}{!}{\includegraphics{300610_mdi.pshalpha.ps}}\hspace*{.5cm}\resizebox{3.cm}{!}{\includegraphics{010710_mdi.pshalpha.ps}}\hspace*{.5cm}\resizebox{3.cm}{!}{\includegraphics{020710_mdi.pshalpha.ps}}\hspace*{.5cm}\resizebox{3.cm}{!}{\includegraphics{120411_mdi.pshalpha.ps}}\\$ $\\
%\resizebox{3.cm}{!}{\includegraphics{040505_mdi.psmagn.ps}}$ $\\

\resizebox{12.cm}{!}{\includegraphics{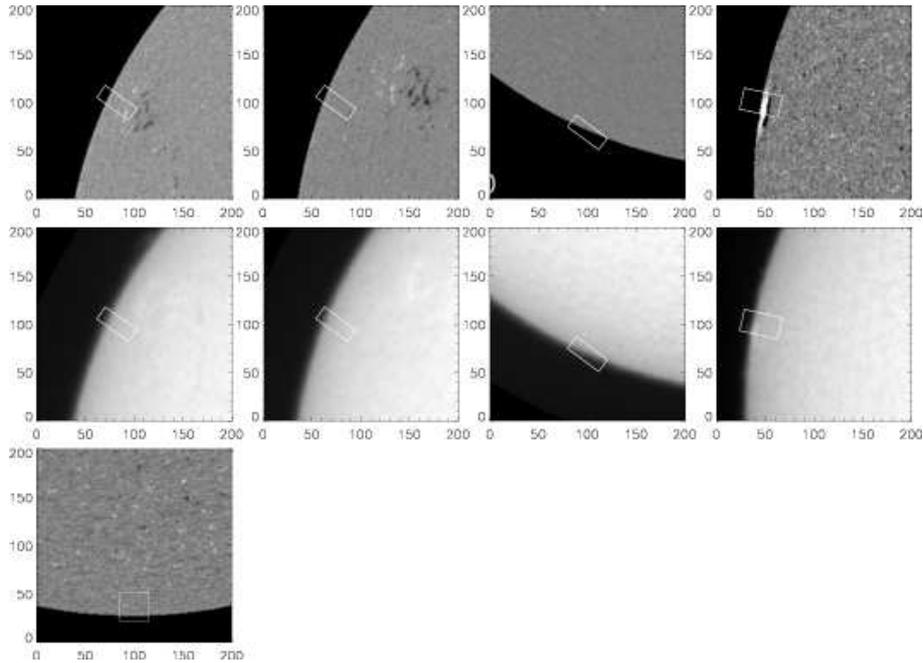}}

\caption{Location of the FOVs on full-disk LOS magnetograms and H$\alpha$
    images. Top row (2nd row), left to right: observations No.~1, 2/3, 4 and 5
    overlaid on a magnetogram (H$\alpha$ image). Bottom row: observation No.~6 overlaid on a magnetogram. Tick marks are in arcsecs. }\label{figlocation}
\end{figure}

Figure \ref{figlocation} shows the locations of the FOVs of the observations overlaid on cut-outs from LOS magnetograms from MDI
  (observations No.~1, 2/3, 4 and 6) and HMI (No.~5) and from GONG H$\alpha$ images (all but No.~6), respectively.  All data apart from observation No.~5 were taken in areas of quiet Sun (QS), while No.~5 was taken in active region (AR) NOAA 11191. There are no discernible large-scale structures such as prominences in the H$\alpha$ images that could have had an impact on the observations close to
  the limb. All observations benefited from the real-time correction of wavefront deformations by the Kiepenheuer-Institut AO system \citep[KAOS;][]{vdluehe+etal2003}. Apart from the observations in 2011, where a sunspot was located close to the limb, a facula was selected as AO lock point. The latter was possible because at the VTT the light level in the AO wave front sensor could be adjusted by a motorized grey wedge and the AO software removed intensity gradients before the correlation of sub-apertures.

%(cf.~Appendix \ref{flat_field})
\section{Data reduction and analysis\label{sec_datared_ana}}
We corrected the spectra from all instruments for the dark current of the CCDs. The removal of flat-field defects had to be adjusted for each spectral line and instrument. In this section, we describe the additional data reduction steps or analysis methods that are particular for the off-limb data.
\subsection{Stray-light correction\label{stray_corr}}
For observations on the solar disc, stray light contaminates the data with about 20\,\% of spurious light \citep[e.g.,][]{beck+etal2011,beck+etal2013d}. For off-limb regions, the intrinsic intensities are rapidly dropping to below 20\,\% of the disc-centre values \citep[e.g.,][]{beck+rezaei2011} and thus stray light becomes critical. There are some observational approaches to minimize stray light in off-limb observations, such as putting the spectrograph slit parallel to the limb \citep{centeno+etal2010} or blocking the part of the slit (or FOV in general) that remains on the disc with an opaque cover similar to a coronagraph \citep{socasnavarro+elmore2005}. However, some contamination of off-limb observations with light from the disc can still not be avoided. 

To reduce the stray-light contamination of the data, we used two different
methods. The first was based on the -- theoretically correct -- approach of
modeling the off-limb stray light from the intensities observed on the disc
\citep{zwaan1965,staveland1970,mattig1983,martinezpillet1992,beck+etal2011}.
A spatially- and/or temporally-averaged line profile was calculated for the
on-disc region of the FOV that was furthest from the limb (about
10$^{\prime\prime}$-30$^{\prime\prime}$) in each data set. Using the observed intensities on cuts across the
limb, or a theoretical modeling or measurement of the spatial point-spread
function (PSF), a suited stray-light fraction as a function of the limb
distance can be determined
\citep[cf.][]{beck+etal2011,loefdahl+scharmer2012}. The fraction is multiplied
with the average on-disc profile and subtracted from the observed
spectra. This method has the drawback that the region of the FOV that is still
on the disc might actually not cover the full area from which the stray light
originated from, e.g., the FOV would always need to cover the full solar disc in theory. This method thus commonly leaves some residuals far away from the limb because the average profile used is not identical to the real stray-light profile, but it provides a smooth stray-light correction across the limb without any discontinuities or intensity jumps at the limb location. 

There exists a second, more empirical approach for off-limb stray-light
correction. The stray-light profile to be subtracted is determined from some
off-limb region that is far away from the limb and from any true solar feature
\citep[e.g.,][]{sanchezandrade+etal2007,martinezgonzalez+etal2012}. With this
approach, one uses the result of the stray-light contamination instead of the
theoretical source of the stray light. The fraction of the stray-light profile
to be subtracted can then be determined from the residual intensity at
continuum wavelengths divided by the corresponding intensity in the
stray-light profile. The advantage of this method is a good stray-light
correction far away from the limb, but it usually shows a worse correction
close to the limb. It also creates a discontinuity in intensities at the location of the limb itself, which also has to be defined by some ad-hoc criterion. Because we are mainly interested in the properties of off-limb spectra, we used the second method for most data sets, apart from those where a stray-light correction based on the first method already had been applied to the data for previous studies. No stray-light correction was applied to the reconstructed GFPI spectra that showed little stray light after the deconvolution.
\subsection{Determination of line parameters}
Most of the data are either spatially or spectrally over-sampled. To increase the signal-to-noise ratio, we thus binned most of the spectra by two in the spatial and/or spectral dimension before deriving the line properties in individual profiles.

For all observations and all wavelength ranges, we used the average line profile from the on-disc region of the FOV that was
furthest from the limb to determine an intensity normalization coefficient
and any other needed corrections, e.g., linear or low-order intensity trends
in the dispersion direction, that matched the average observed spectra to the
corresponding Fourier transform spectrometer
\citep[FTS;][]{kurucz+etal1984,neckel1999} solar atlas profiles. All other
spectra of a given data set were then normalized and/or corrected with these
values. For all on-disc and off-limb spectra, we derived the continuum
intensity $I_c$ from some continuum wavelength range and a generic line-core
intensity $I_{\rm core}$ from an integration over the line-core region in the
spectra. The maps of $I_c$ were used to co-align the observations in different
wavelengths that showed offsets by differential refraction \citep[e.g.,
Appendix A of][]{beck+etal2008,felipe+etal2010}.

To determine the line width in individual spectra, we fitted single Gaussians to the
line profiles. The central self-absorption in the line profiles disappears
for heights larger than about 3\,Mm above the limb, leaving a roughly Gaussian
shape (see Figs.~\ref{av_prof_1} and \ref{av_2} below).
%\textbf{\textcolor{red}{([DF: Should we have tried to fit any other profile, at least for the GFPI data (which, as said two paragraphs below, are not fitted so well)? Any more discussion in the text here or elsewhere about choice of Gaussian vs. other fitting?])}}
The spectral range to be analyzed was restricted each time to cover only the respective emission profile of a single line for the fit. The \ion{Ca}{ii} H spectra recorded with a PCO were the only case where two separate fits were performed to a single profile, i.e., one for \ion{Ca}{ii} H and one for H$\epsilon$. The Gaussian fit yielded the central amplitude, the central position and the full-width at half-maximum (FWHM) of the Gaussian. The position of the Gaussian was converted to the corresponding LOS velocity using the line-core position in the average profiles as zero-point reference. 

As a cross-check of the width derived by the Gaussian fit, we also used a simpler method that only determined the maximal intensity of the emission profile and the two positions where the intensity dropped to half of it. The FWHM is then directly given by the distance between these two positions.

The fit of a Gaussian function to the GFPI spectra worked less reliably because the outermost wavelength points in the blue and red wings often still sample the line emission without a clear drop of intensities, which also made the determination of the FWHM with the direct method impossible. The line parameters for the GFPI spectra are thus not as well defined as for the other data.
\subsection{Derivation of kinetic temperature and non-thermal line width}
The line width $\Delta\lambda$ of emission lines formed in the optically-thin regime can be described by
\begin{eqnarray}
\Delta \lambda = \frac{\lambda}{c}\sqrt{\frac{2\,RT_{\rm kin}}{\mu}+v^2_{\rm non-th}} \,,\label{eq1}
\end{eqnarray}
where $\lambda$ is the central wavelength, $c$ the speed of light, $R$ the universal gas constant, $T_{\rm kin}$ the kinetic temperature, $\mu$ the molecular weight and $v_{\rm non-th}$ any additional non-thermal line broadening \citep[e.g.,][]{tandberghanssen1960,bendlin+etal1988}. Magnetic broadening also contributes to the line width, but for typical chromospheric field strengths below 100\,G \citep[e.g.,][]{centeno+etal2010} it presumably is negligible in spicules and prominences.

The line width $\Delta\lambda$ is related to the FWHM in the case of a Gaussian emission profile by 
\begin{eqnarray}
\Delta\lambda = \frac{\rm FWHM}{2\,\sqrt{\ln 2}} = \sqrt{2}\,\sigma \,, \label{eq2}
\end{eqnarray}
where $\sigma$ is the width of the Gaussian\footnote{We note that in some older literature, the use of $\Delta\lambda$ in this context varies, sometimes standing for the half-width at half-maximum (i.~e., $\Delta\lambda = \frac{\rm FWHM}{2}$), while in yet other instances the term ``half-width'' is used for the FWHM or an exponential of the form $\exp^{-(x/\sigma)^2}$ is used.}.

With line widths of two simultaneously observed spectral lines from two different chemical elements with different molecular weights $\mu_i$, one can derive an estimate for both the kinetic temperature and the velocity equivalent of the non-thermal line broadening. For $n > 2$ observed spectral lines, Eq.~\ref{eq1} leads to a set of $n$ equations for the two open parameters $T_{\rm kin}$ and $v_{\rm non-th}$ that can be solved by a matrix inversion or a least-square fit. 
%n over-determined

The whole approach assumes emission profiles of Gaussian shape, i.e., without
central reversals by self-absorption, and that the observed spectral lines
have to form in the same solar atmospheric volume or at least under the same
atmospheric conditions. With the variation of opacity from line to line in the
chromosphere, the second condition needs not be fulfilled automatically even
for features that appear to be on the same spatial location in the data \citep[e.g.,][]{stellmacher+wiehr2015}. The first condition can be estimated from the line shape. 

We determined $T_{\rm kin}$ and $v_{\rm non-th}$ using Eqs.~\ref{eq1} and \ref{eq2} for all pairs of lines and all lines together only for the data taken in 2011 because they could be aligned precisely thanks to their high spatial resolution. The width  $\sigma$ of the Gaussians that were fitted to the profiles was corrected beforehand for the instrumental broadening $\sigma_{\rm instr}$ of the respective spectrograph \citep[][his Eq.~II.12]{tandberghanssen1960}. The value of $\sigma_{\rm instr}$ was determined by convolving the FTS atlas profiles with a Gaussian to match the line width of photospheric lines in the observed spectra \citep[cf.][]{cabrerasolana+etal2007}. The velocity equivalents of the instrumental broadenings were, however, only about 2\,km\,s$^{-1}$, which turned out to be nearly negligible in comparison to $v_{\rm non-th}$. Therefore, this correction had only a small impact on the final values of $T_{\rm kin}$ and $v_{\rm non-th}$.
\section{Results\label{sec_results}}
\begin{figure}
%\hspace*{1cm}\resizebox{11.cm}{!}{\includegraphics{gfpi_exam_new.ps}}$ $\\$ $\\
\resizebox{12.cm}{!}{\includegraphics{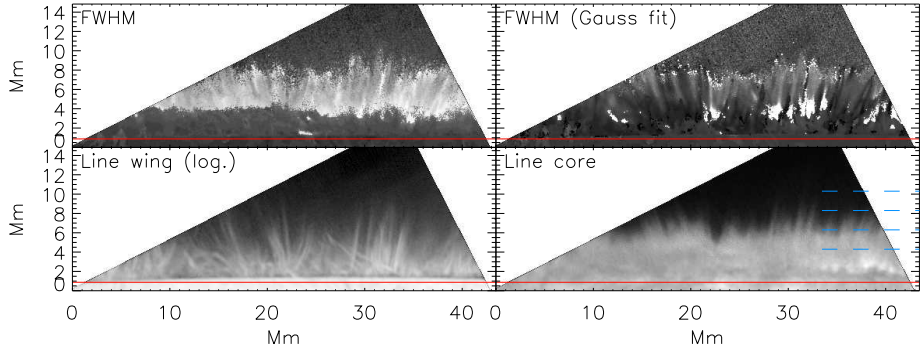}}
\caption{Example of the GFPI H$\alpha$ observations at the limb (observation
  No.~6). Bottom row, left to right: line-wing intensity in logarithmic
  display and line-core intensity. Top row, left to right: FWHM from direct
  determination and FWHM from fit of a Gaussian. The red horizontal line marks
  the approximate location of the limb. The blue dashed lines in the
  lower-right panel indicate the heights above the limb of the spectra shown
  in Fig.~\ref{gfpi_temp}.  An animation of the complete time-series is
  available in the online section.}\label{gfpi_exam} %Note the difference between the line-wing and line-core images: individual spicules seen in the line-wing image have almost no correspondence in the line-core image.
\end{figure}
\begin{figure}
%\resizebox{6.cm}{!}{\includegraphics{gfpi_spic_exam_wing.ps}}\hspace*{.15cm}\resizebox{6.cm}{!}{\includegraphics{gfpi_spic_exam_width_gauss.ps}}$ $\\$ $\\ 
\resizebox{12.cm}{!}{\includegraphics{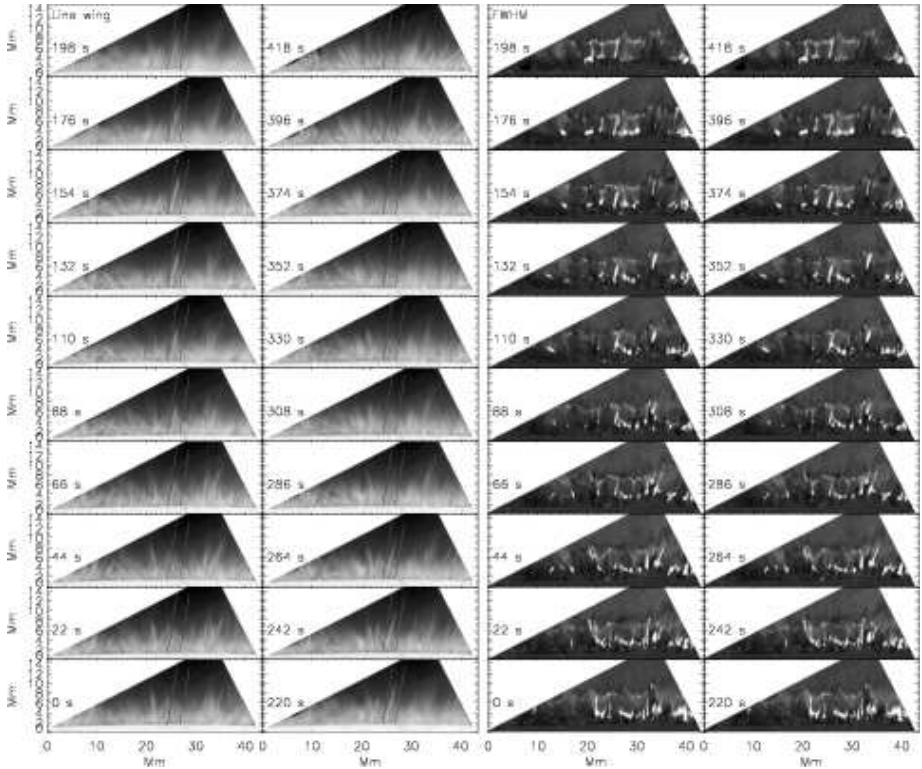}}
\caption{Evolution of an individual spicule in line-wing images (left panels) and in maps of the FWHM from a Gaussian fit (right panels) in the GFPI H$\alpha$ data. The spicule is located between the two inclined red lines in each sub-panel. It starts at about $t= 44$\,s and has disappeared at $t= 396$\,s. See also the animation of the complete time-series in the online section.} \label{spic_evol_gfpi}
\end{figure}
\subsection{2D spectroscopy in H$\alpha$}
\subsubsection{GFPI data}
Figure \ref{gfpi_exam} shows one of the 39 scans through the H$\alpha$ line taken with the GFPI (observation No.~6, setup 3). An animation of the complete time-series is available in the online section \citep[see also][]{puschmann2016}. Individual spicules can best be identified in the line-wing image, while in the line-core image the complete off-limb region shows a more diffuse emission pattern. In the maps of the FWHM from the Gaussian fit (top right panel of Fig.~\ref{gfpi_exam}), individual spicules show an increased FWHM relative to their surroundings, but in most cases without any clear trend along their length.

\begin{figure}
%\resizebox{11.cm}{!}{\hspace*{1cm}\includegraphics{gfpi_spic_exam.ps}}$ $\\$ $\\
\resizebox{12.cm}{!}{\includegraphics{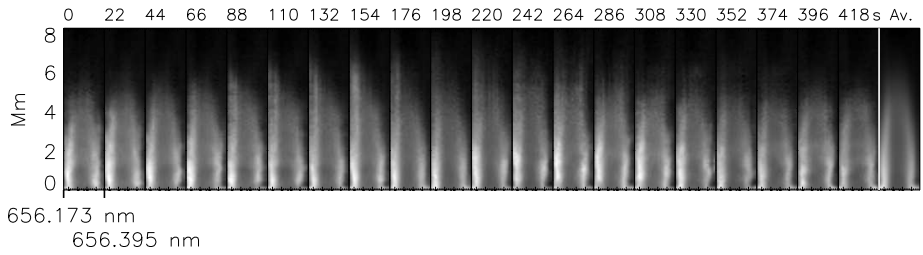}}
\caption{Individual spectra during the evolution of the spicule indicated in Fig.~\ref{spic_evol_gfpi}. The wavelength increases from left to right in each sub-panel, within the range marked for the leftmost sub-panel. The rightmost panel shows the off-limb spectra in the GFPI data averaged over all 39 bursts and all spatial positions along the limb.} \label{spec_temp_gfpi}
\end{figure}

To trace the temporal evolution of the emission, we selected one prominent spicule within the time-series (Figs.~\ref{spic_evol_gfpi} and \ref{spec_temp_gfpi}). It can be clearly distinguished in the line-wing images at $t=44\,$s (left column in Fig.~\ref{spic_evol_gfpi}, third panel from the bottom) and can be followed until $t=396\,$s. Its line width is larger than that of the surroundings (Fig.~\ref{spic_evol_gfpi}, third and fourth column), but reduces with increasing limb distance in most cases. The spectra along the central axis of the spicule (Fig.~\ref{spec_temp_gfpi}) resemble those from the TESOS data in Fig.~\ref{ha_av}, but show larger Doppler shifts, especially towards the blue at the beginning ($t=88 - 220$\,s). The spicule expands up to a height of about 6\,Mm at $t=242\,$s. The average spectra from all spatial locations and all of the 39 wavelength scans show a reduction of the line width with increasing limb distance (rightmost panel in Fig.~\ref{spec_temp_gfpi}).

\begin{figure}
%\begin{minipage}{6cm}
%\resizebox{5.cm}{!}{\includegraphics{gfpi_temp1.ps}}$ $\\$ $\\
%\resizebox{5.cm}{!}{\includegraphics{gfpi_temp2.ps}}
%\end{minipage}
%\begin{minipage}{6cm}
%\resizebox{5.cm}{!}{\includegraphics{gfpi_temp3.ps}}$ $\\$ $\\
%\resizebox{5.cm}{!}{\includegraphics{gfpi_temp4.ps}}
%\end{minipage}$ $\\$ $\\
\resizebox{12.cm}{!}{\includegraphics{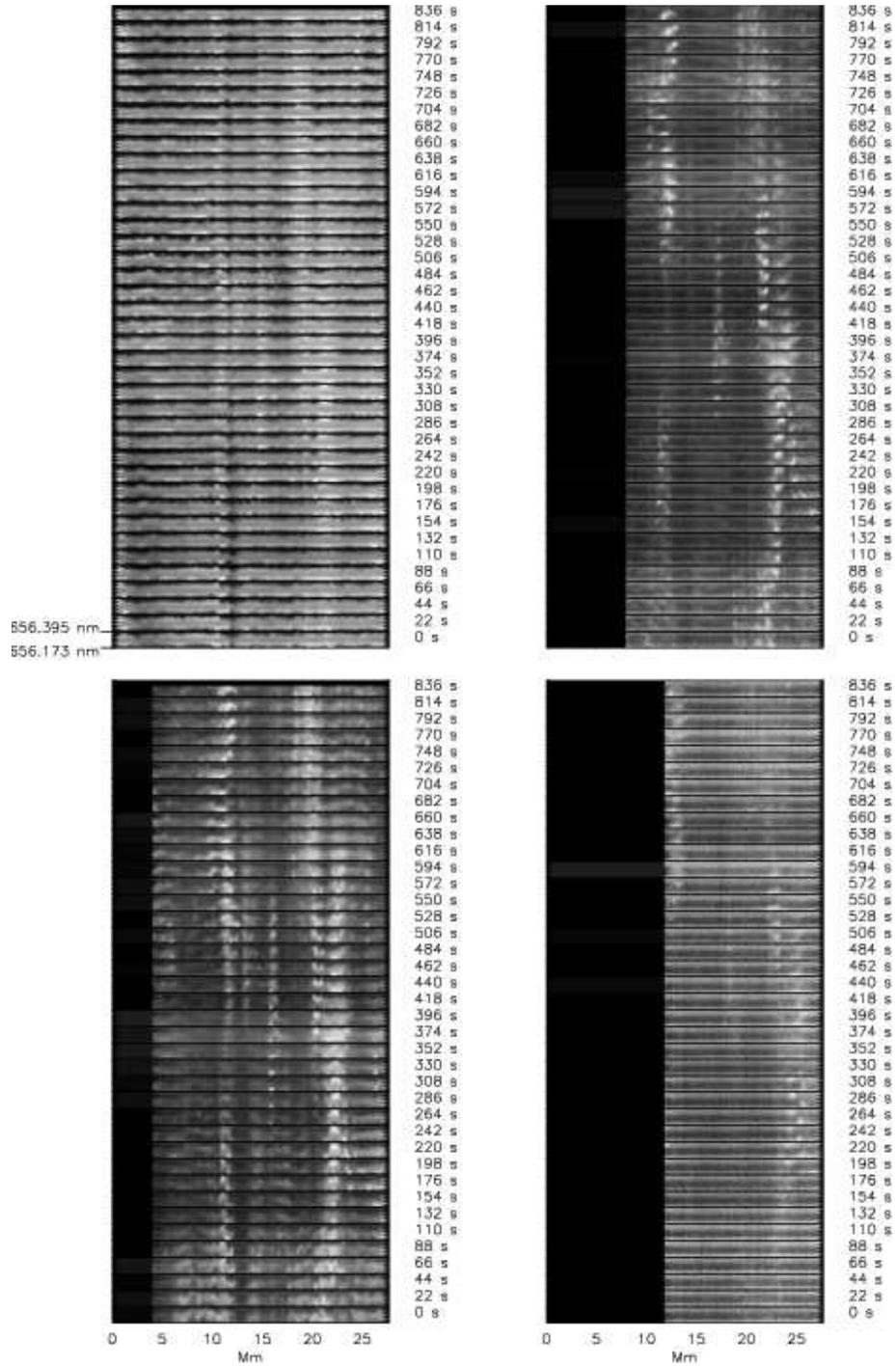}}
\caption{Temporal evolution of H$\alpha$ spectra at different heights above the limb in the GFPI data. Left column: $h= 4.3$\,Mm (top),  $h= 6.3$\,Mm (bottom). Right column: $h= 8.3$\,Mm (top),  $h= 10.3$\,Mm (bottom). Time increases from bottom to top and the spatial position along the limb increases from left to right in each panel. The wavelength increases from bottom to top in each sub-panel, within the range marked in the lower sub-panel of the top-left panel. See also the animation of the complete time-series in the online section.}\label{gfpi_temp}
\end{figure}

Because of the difficulties in extracting the line width from the GPFI
spectra, we also selected individual spectra at different heights above the
limb to follow their temporal evolution (Fig.~\ref{gfpi_temp} and the
animation in the online section). Figure \ref{gfpi_temp} shows the temporal
evolution of individual spectra at four of those heights above the limb. Like
in the animation, the presence of large and varying Doppler shifts is clearly
visible, but the disappearance of some bright emission is usually not
connected to an increase in line width but only to a fading with time.

\begin{figure*}
\resizebox{12.cm}{!}{\includegraphics{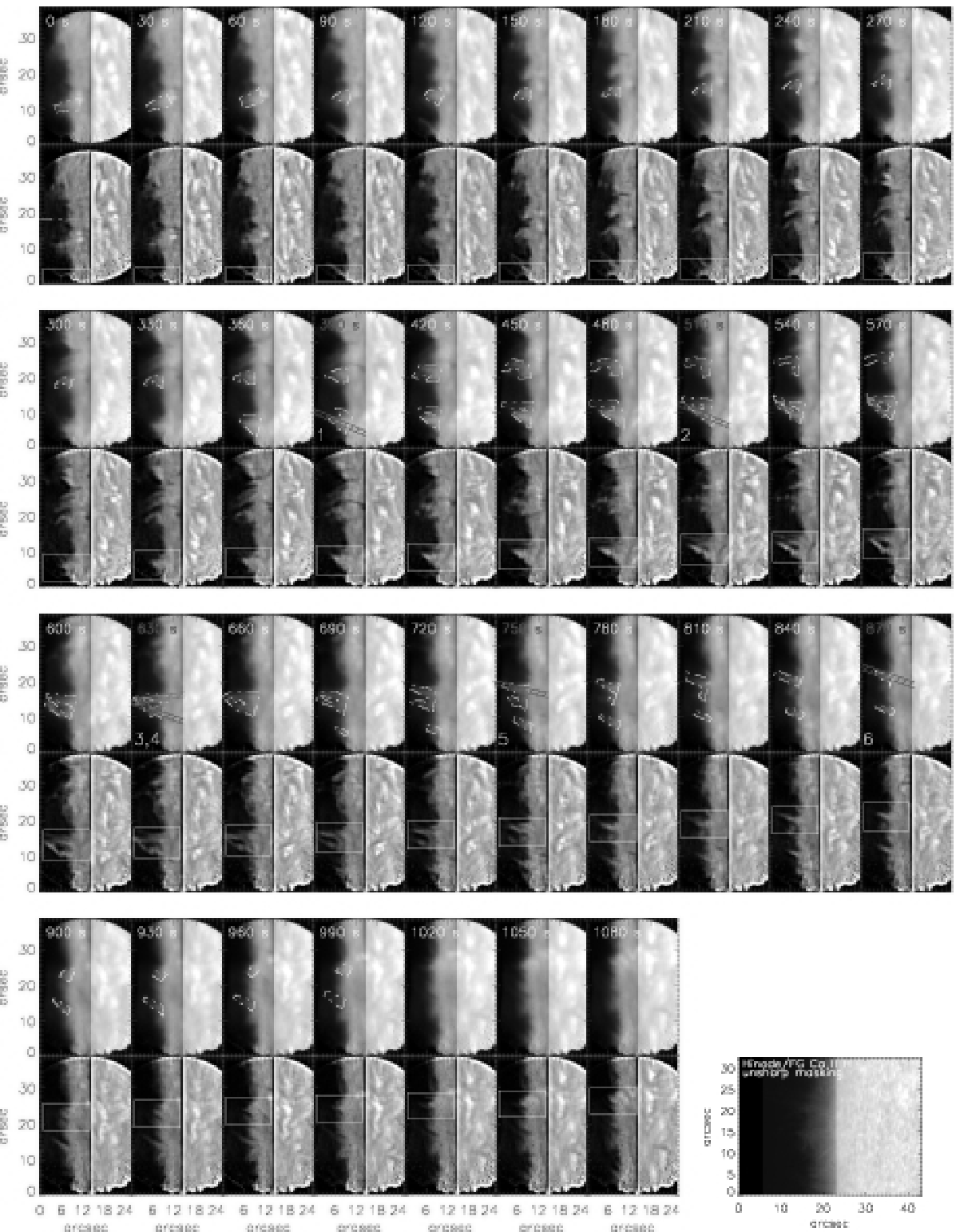}}
\caption{H$\alpha$ line-core images acquired with TESOS during observation
  No.~1. Time increases from left to right and top to bottom. For each time
  step, the original (unsharp-masked) line-core image is displayed in the
  upper (lower) panel. The white dash-dotted contour lines mark the area of
  some individual spicules. The white rectangles follow the location of one specific feature
  with time. The red- and blue-hatched areas (with corresponding time of
  observation given in red in the relevant sub-panels) denote the range of
  averaging for the profiles 1 to 6 shown in Fig.~\ref{ha_av}. The orange
  dash-dotted line at $t=0$\,s indicates the location of the TIP slit. The
  bottom rightmost panel shows a cut-out from a broad-band Ca II H image from
  Hinode for comparison.}\label{fig_ha}
% \textbf{\textcolor{red}{([DF: ] Long. Explanation in captions should be
% avoided, only short descriptive sentence of what figures globally
% show. Then, details in text (already given).)}}The spatial scanning by TIP moves the H$\alpha$ FOV along the $y$-axis.
\end{figure*}
\begin{figure}
%\centerline{\resizebox{8.8cm}{!}{\hspace*{1cm}\includegraphics{halpha_offlimb_cuts.ps}}}$ $\\
\centerline{\resizebox{10.cm}{!}{\includegraphics{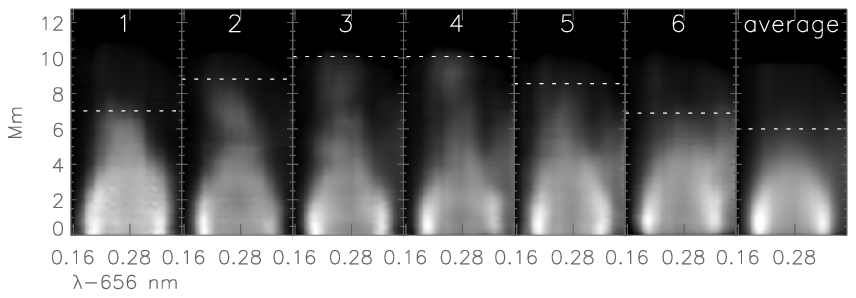}}}
\caption{Average H$\alpha$ profiles in one spicule feature. {Left to right/1\,--\,6}: red hatched area of Fig.~\ref{fig_ha} at $t\,=\,390, 510, 630$\,s, blue hatched area at $t\,=\,630$\,s, red hatched area at 750 and 870\,s. The rightmost panel shows the average off-limb profiles. The horizontal white dotted lines denote the maximal height where the spectra are still significant. } \label{ha_av}
\end{figure}
\subsubsection{TESOS data}
Figure \ref{fig_ha} shows line-core images in H$\alpha$ taken with TESOS during observation No.~1 (cf.~Table \ref{obs_specs}, setup 1). The cadence between subsequent images is about 30\,seconds. The corresponding FOV in the other spectral lines is shown in Fig.~\ref{all_wl} below. In the top left panel of Fig.~\ref{fig_ha}, the approximate location of the TIP slit inside the TESOS-FOV is indicated by a dash-dotted horizontal line. Each line-core image is displayed twice: with and without an unsharp masking that enhances small-scale features. The limb location was determined from the corresponding continuum intensity images (not shown) and was used to delineate the region in which the stray-light correction with the second correction method (Sect.~\ref{stray_corr}) was applied. Its approximate location shows up at about the middle of the FOV as a vertical line in each image. The FOV is slowly drifting with time because of the co-temporal scanning of the slit-spectrograph instruments. To extract the spectral properties of individual spicules, we masked some intensity enhancements in the H$\alpha$ line core above a height of about 5\,Mm beyond the limb (white dash-dotted contours in the top panels of a given row). We note that below this height it is nearly impossible to reliably identify individual features in line-core images because of the complete overlap of different features. %, even if the FOV there has the same spatial resolution as the rest of the image,
% -- postponing any possible classification of its type -

We selected one feature  that was covered during all of its life-time (cf.~the
white rectangles in Fig.~\ref{fig_ha}) to determine laterally averaged
profiles along its extent at five different time steps (cf.~the six blue and
red-hatched areas marked in Figure \ref{fig_ha}). The feature, or at least a
predecessor of the feature at the same spatial location, can be identified
already at $t\,=\,60$\,s (third column in top row; $x,y\sim 8^{\prime\prime}, 3^{\prime\prime}$) if one follows its evolution backwards in time. It rises over the course of $\sim 300$ s to a maximal height of about 16$^{\prime\prime}$ above the limb at $t\,=\,630$\,s. It retracts subsequently towards the limb and cannot be identified anymore at $t\,=\,930$\,s. 

Figure \ref{ha_av} shows the spatially-averaged line profiles, i.~e., those
obtained averaging laterally along the $y$-axis of the images in
Fig.~\ref{fig_ha} over the colored hatched areas (labeled 1 to 6 in Figure
\ref{ha_av}) at the corresponding time step during the observations, with limb
distance increasing along the vertical direction in the figure. The feature is
roughly oriented (tilt of less than 30 degrees) perpendicular to the limb, so
the laterally averaged profiles roughly sample its extent in height above the
limb. The temporal variation of its maximal height (marked as a horizontal
white dotted line in each panel of Fig.~\ref{ha_av}) is roughly parabolic. The profiles all show a change from self-absorption with a central absorption core up to a height of about 2\,Mm above the limb to pure emission profiles at higher layers. In the visual impression, the line width is constantly reducing with height. The same holds for the average off-limb profiles shown in the rightmost panel of Fig.~\ref{ha_av} that were retrieved by averaging over nearly the full extent of the FOV, excluding the top and bottom borders because of the curved field stop. In the profiles averaged over the full usable FOV, only spectra up to about 6\,Mm above the limb have intensities higher than the noise level, similar to the GFPI spectra in Fig.~\ref{spec_temp_gfpi}. The profiles marked as 4 to 6 in Figure \ref{ha_av} exhibit a faint haze at the red end of the spectrum ($\lambda\sim 656.34$\,nm) which is some flat-field residual that could not be corrected for.
\begin{figure}
%\centerline{\resizebox{4.cm}{!}{\includegraphics{halpha_cut3.ps}}\hspace*{-.5cm}\resizebox{4.cm}{!}{\includegraphics{halpha_cut3a.ps}}}
\centerline{\resizebox{8cm}{!}{\includegraphics{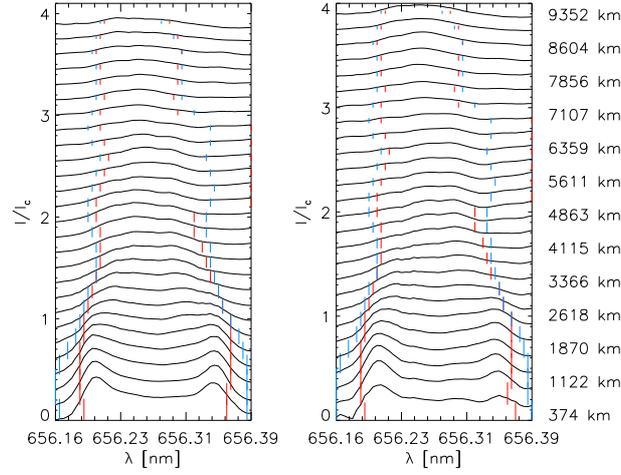}}}
\caption{Individual profiles from the laterally averaged sets of spectra Nos.~3 (left) and 4 (right) of Fig.~\ref{ha_av}. The profiles are displaced from each other in $y$ for better visibility. The corresponding heights above the limb are given at the right-hand side. The blue/red bars denote the FWHM from a Gaussian fit and the locations where the intensity drops to 50\,\% of its maximum, respectively.}\label{indiv_spec}
\end{figure}
\begin{figure}
%\centerline{\resizebox{12.cm}{!}{\includegraphics{halpha_linewidth_cuts_new.ps}}}
\centerline{\resizebox{12cm}{!}{\includegraphics{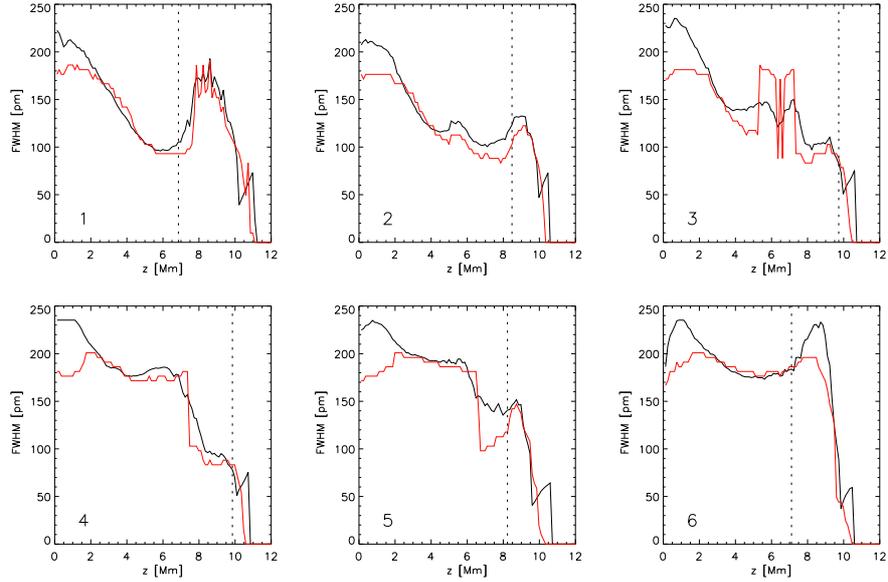}}}
\caption{The FWHM of the same  H$\alpha$ profiles of Fig.~\ref{ha_av}. {Black}: FWHM from the Gaussian fit. {Red}: FWHM from the locations where the intensity drops to 50\,\% of the maximum. The vertical dotted lines denote the maximal height where the spectra are still significant. }\label{fwhm_ha}
\end{figure}

The individual profiles along the two branches of the feature that have
developed at $t=\,$\,630\,s (marked as red and blue-hatched areas in the
corresponding panel of Fig.~\ref{fig_ha}) are shown in
Fig.~\ref{indiv_spec}. The FWHM derived from the Gaussian fit and that derived
directly are over-plotted with short red and blue coloured bars,
respectively. The FWHM is seen to reduce up to a height of about 5\,Mm above
the limb, and remains at best constant at higher layers. The corresponding
FWHMs in all six sets of laterally averaged profiles are shown in
Fig.~\ref{fwhm_ha}. Up to the height where the intensity in the spectra is
still significant, all six examples of the height variation in the FWHM in
resolved features show a monotonic decrease in width. The FWHM only increases
just where the spectra level off into the noise (profile sets 1, 2 and 6 in
Fig.~\ref{fwhm_ha}). The FWHM retrieved via Gaussian fitting agrees reasonably
well for most spectra with that obtained using the direct method, which,
however, seemingly tends to give noisier results. This might be related to the
fact that the Gaussian fit makes use of all wavelength points in the line
profile, whereas the direct method is based on three wavelength points only,
i.e., the one corresponding to maximum intensity and the two locations for which the line profile reaches a level of 50\,\% of that value. The latter method is therefore more sensitive to noise peaks. 

Other examples of time-series of H$\alpha$ line-core images are shown in
Figs.~\ref{ha_1} to \ref{ha_4} in Appendix \ref{app_ha}. They all share more
prominently than Fig.~\ref{fig_ha} a certain peculiarity, which, however, is
also apparent in the latter: the majority of the individual structures or
features that are seen above a limiting height of about 5 to 6\,Mm show little
to no resemblance to spicules, i.e., only a very few features are elongated, slender brightenings jutting out and away from the limb. Figures~\ref{ha_1} and \ref{ha_2} clearly show material, or more precisely, brightenings that recede from the limb upwards (bottom rows of both figures), but the corresponding shapes are either elongated structures that are parallel to the limb, or roundish blobs. The same holds for Fig.~\ref{ha_3}: only complex-shaped features appear at heights of more than about 6\,Mm above the limb in the H$\alpha$ line-core images.

\begin{figure}
%\centerline{\begin{minipage}{7cm}\vspace*{.2cm}\hspace*{1cm}\resizebox{6.cm}{!}{\includegraphics{FOV_300610.ps}}$ $\\$ $\\\end{minipage}\hspace*{1cm}\begin{minipage}{5cm}\resizebox{3.3cm}{!}{\includegraphics{hinode_obs_raw_ca_limb.ps}}\\\resizebox{3.3cm}{!}{\includegraphics{hinode_obs_ca_limb_new.ps}}\vspace*{.2cm}\\\end{minipage}}
\centerline{\resizebox{12cm}{!}{\includegraphics{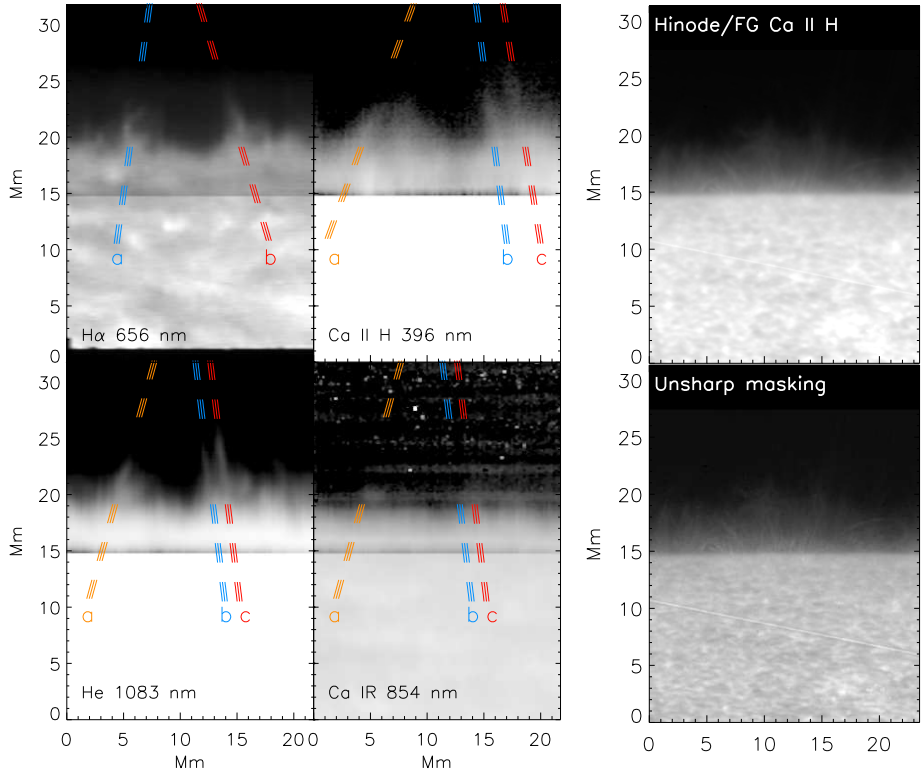}}}
\caption{Left: FOV of observation No.~1 in line-core images of different
  spectral lines. Clockwise, starting left top: artificial H$\alpha$ scan,
  \ion{Ca}{ii} H, \ion{Ca}{ii} IR at 854.2 nm, and \ion{He}{i} at 1083 nm. The
  inclined colored lines labeled {\em a,b,c} denote the regions over which the profiles shown in Fig.~\ref{spic_allwl} were laterally averaged. The two panels at the right-hand side show an equal-sized region from a broad-band Ca II H image from Hinode without (top) and with an unsharp masking (bottom) for comparison.}\label{all_wl}
\end{figure}
\subsection{(Pseudo-)slit-spectrograph data}
The left four panels of Fig.~\ref{all_wl} show the corresponding FOV of observation No.~1 in the slit-spectrograph data. We constructed a pseudo-scan map \citep[e.g.,][]{beck+etal2007} of the same FOV from the 2D H$\alpha$ spectra from TESOS. We selected the TESOS wavelength scan closest in time to each scan step of the slit-spectrograph instruments for that purpose and cut out the corresponding spectra along the location of the TIP slit from the TESOS 2D FOV (top left panel of Fig.~\ref{fig_ha}). The resulting map (upper left subpanel of Fig.~\ref{all_wl}) shows how the temporal evolution of Fig.~\ref{fig_ha} is sampled in a slit-spectrograph observation. A prominent change is that most laterally extended structures in the 2D H$\alpha$ line-core images appear as a series of resolved, rather thin and elongated features in the pseudo-scan map. We did not try to improve the spatial alignment because due to the large differential refraction at the early time of this observation and the sequential scanning, all other spectra in the different wavelength ranges are not strictly co-spatial and simultaneous anyway. 

The line-core images in H$\alpha$, \ion{Ca}{ii} H and \ion{He}{i} at 1083\,nm
are to some extent similar, showing one set of
spicular features at $x\sim 5$\,Mm, a region of reduced extent of  emission
from $x\,=\,$7 to 12\,Mm, and a double pair of spicules at $x\sim
14$\,Mm. These three lines should therefore form in a similar volume, i.e.,
their optical depth should be comparable and they should sample the same
atmospheric volume. The line-core image of \ion{Ca}{ii} IR at 854.2\,nm
differs significantly from all others. The extent of emission is limited to
about 5\,Mm above the limb only and little to no isolated features can be
identified. The two panels at the right-hand side of Fig.~\ref{all_wl} show an equally-sized region from a broad-band \ion{Ca}{ii} H image from the Hinode filtergraph for comparison. For the bottom panel, the image was treated with an unsharp masking to enhance the contrast. The Hinode Ca image shows how the seemingly uniform emission from the limb to about 5\,Mm height above it in the line-core images is composed of individual strands of spicules. The only clear examples of isolated, individual spicules in the spectra (cf.~the inclined coloured lines) are found at heights where the Hinode Ca image exhibits little to no emission anymore.
\begin{figure}
%\begin{center}\hspace*{.8cm}\begin{minipage}{8.cm}\resizebox{3.cm}{!}{\includegraphics{ha_2dcut.ps}}\hspace*{1.5cm}\resizebox{3cm}{!}{\includegraphics{caiih_2dcut.ps}}\vspace*{.5cm}\\\resizebox{4.cm}{!}{\includegraphics{tip_2dcut.ps}}\hspace*{1.55cm}\resizebox{2.cm}{!}{\includegraphics{cair_2dcut.ps}}\\$ $\\\end{minipage}\end{center}
\centerline{\resizebox{9cm}{!}{\includegraphics{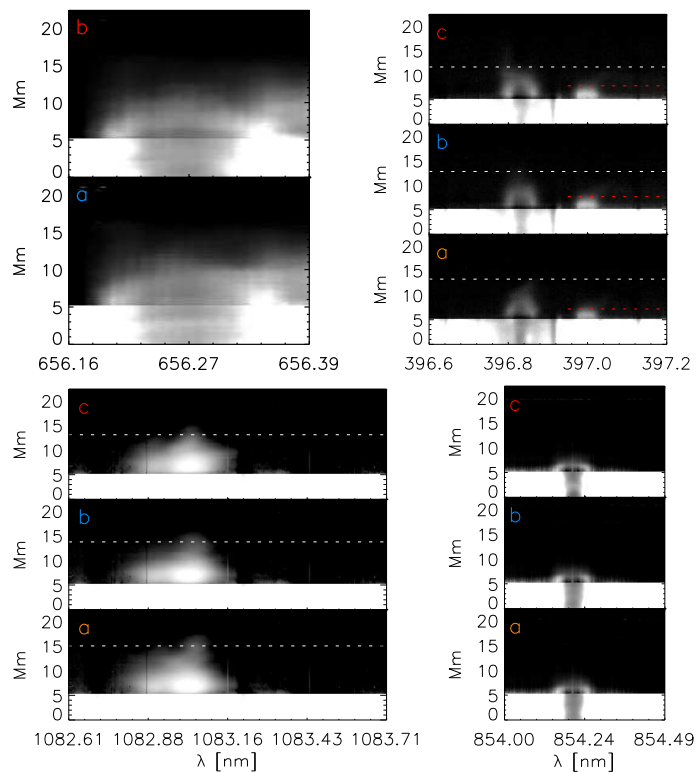}}}
\caption{Laterally averaged spectra along the colored lines labeled {\em a,b,c} in Fig.~\ref{all_wl}. {Clockwise, starting left top}: H$\alpha$, \ion{Ca}{ii} H, \ion{Ca}{ii} IR at 854.2 nm, and \ion{He}{i} at 1083 nm. The horizontal dotted lines denote the maximal height with spectra above the noise level.}\label{spic_allwl}
\end{figure}

\begin{figure}
%\hspace*{-1cm}\begin{minipage}{14cm}\resizebox{4.2cm}{!}{\includegraphics{tip_cut1.ps}}\hspace*{-.5cm}\resizebox{4.2cm}{!}{\includegraphics{tip_cut2.ps}}\hspace*{-.5cm}\resizebox{4.2cm}{!}{\includegraphics{tip_cut3.ps}}\vspace*{-.75cm}\\\resizebox{4.2cm}{!}{\includegraphics{caiih_cut1.ps}}\hspace*{-.5cm}\resizebox{4.2cm}{!}{\includegraphics{caiih_cut2.ps}}\hspace*{-.5cm}\resizebox{4.2cm}{!}{\includegraphics{caiih_cut3.ps}}\end{minipage}
\centerline{\resizebox{12cm}{!}{\includegraphics{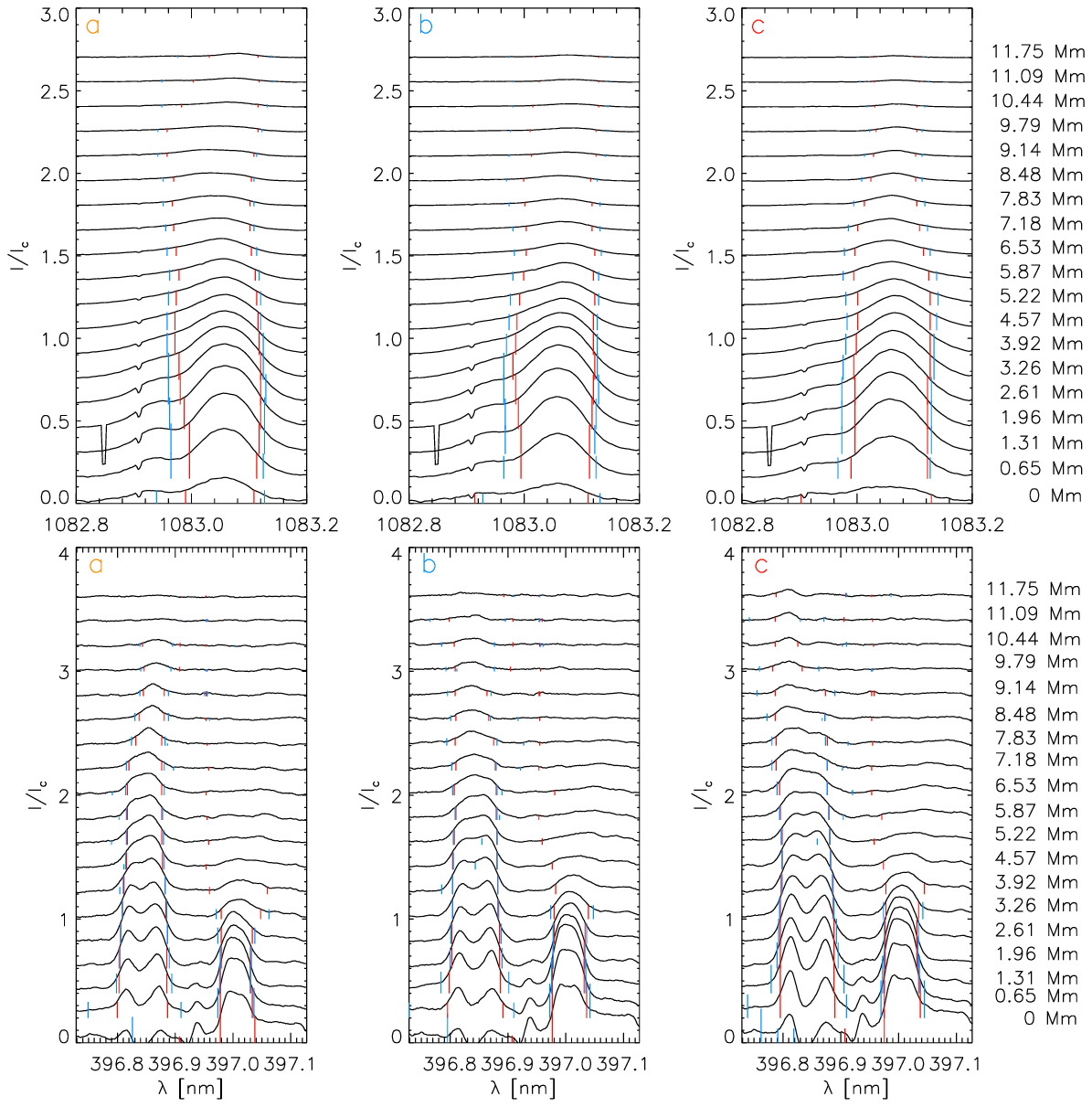}}}
\caption{Laterally averaged spectra along the colored lines labeled {\em a,b,c} in Fig.~\ref{all_wl}. {Top row}: \ion{He}{i} at 1083\,nm. {Bottom row}: \ion{Ca}{ii} H and H$\epsilon$.  The blue/red bars denote the FWHM from a Gaussian fit and the locations where the intensity drops to 50\,\% of the maximum, respectively.} \label{indi_allwl}
\end{figure}

To quantify the properties of the three (two in H$\alpha$) spicules in the
different spectral lines, we again averaged the spectra laterally, i.e., along
the $x$-axis, over the extent of the inclined coloured lines in
Fig.~\ref{all_wl}. As before, the spicules are nearly perpendicular to the
limb, so the variation with limb distance samples the variation along the
spicule length at the same time. The corresponding sets of profiles are shown
in Fig.~\ref{spic_allwl}. All spectra are displayed on a logarithmic intensity
scale to enhance the visibility of their shapes. The profile shape and its
variation with height for H$\alpha$ and the \ion{Ca}{ii} H and IR lines is
similar. The spectra show a self-absorption core close to the limb that
changes a single central emission, and a decrease in line width with height \citep[see also][]{pasachoff1970,pasachoff+zirin1971}. For \ion{Ca}{ii} IR at 854.2\,nm, the height scale of this variation is compressed by about two compared to the other lines. The H$\epsilon$ line to the red of the \ion{Ca}{ii} H line core at 397\,nm shows emission to about the same height as \ion{Ca}{ii} IR at 854.2\,nm. The amplitude of emission in H$\epsilon$ exceeds the one of \ion{Ca}{ii} H close to the limb. The shape of the \ion{He}{i} line at 1083\,nm is different from all others because of its specific formation and its being made of several components \citep[e.g.,][]{sanchezandrade+etal2007}, but it shows a similar trend of a reduction in line width with height.

\begin{figure}
%\resizebox{12cm}{!}{\includegraphics{linewidth_all_cuts.ps}}
\centerline{\resizebox{12cm}{!}{\includegraphics{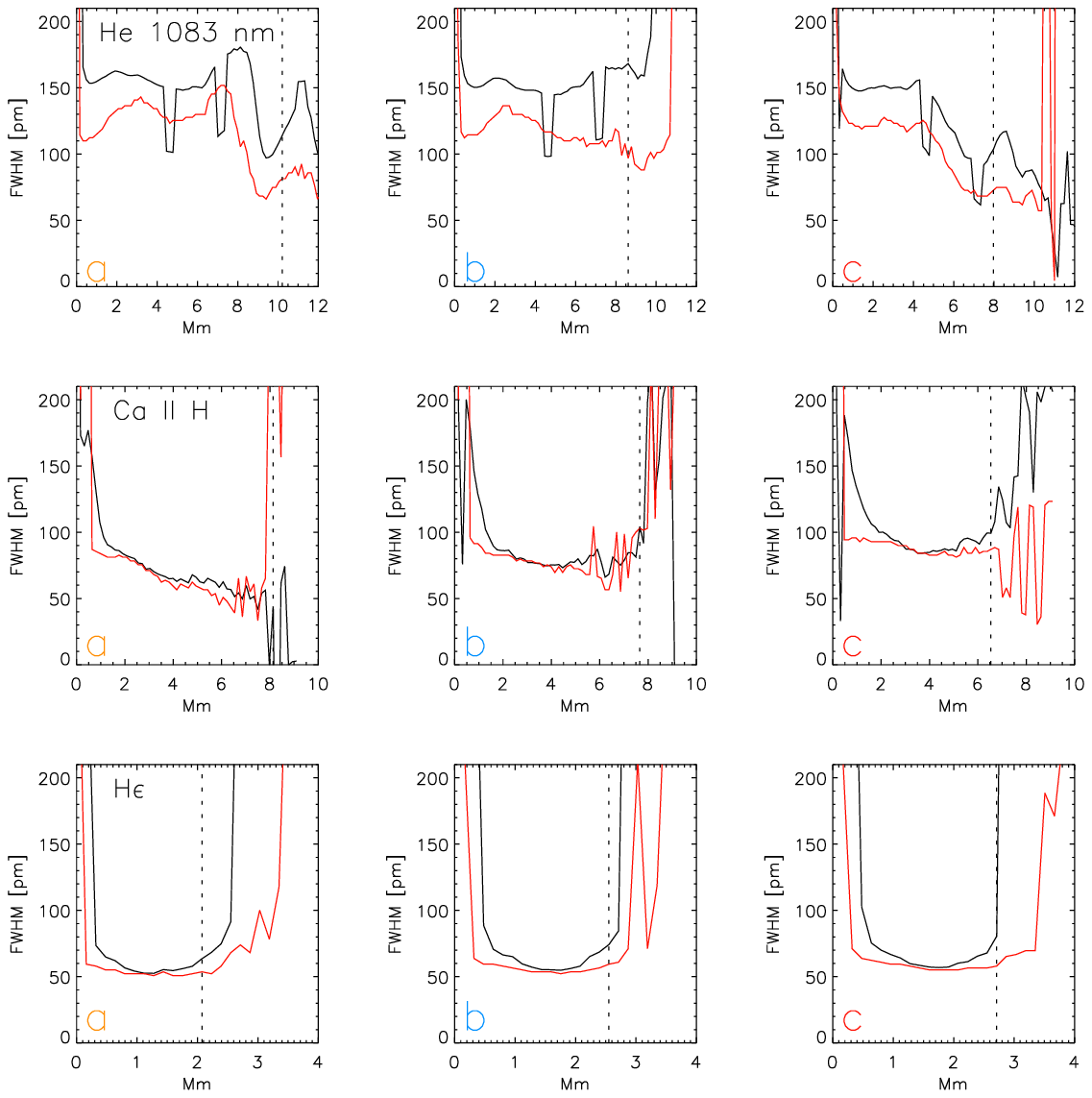}}}
\caption{The FWHM of the laterally averaged spectra along the colored lines
  labeled {\em a,b,c} in Fig.~\ref{all_wl}. {Black}: FWHM from the
  Gaussian fit. {Red}: FWHM from the locations where the intensity drops
  to 50\,\% of the maximum. The vertical dotted lines denote the maximal range
  with spectra above the noise level as marked in Fig.~\ref{spic_allwl}. }\label{fwhm_allwl}
\end{figure}

The individual spectra of \ion{Ca}{ii} H and \ion{He}{i} at 1083\,nm for the three chosen spicules are shown in Fig.~\ref{indi_allwl}. For \ion{He}{i} at 1083\,nm, the line width increases up to a height of about 6\,Mm, but this broadening is artificial and caused by the merging of the red and blue components of the lines. The line width decreases at larger heights. For the \ion{Ca}{ii} H spectra, the line width reduces only slightly up to about 7 to 8\,Mm in all panels, but above that height the emission peak becomes significantly smaller. H$\epsilon$ shows a different behavior, with a broadening of the line at about 3\,Mm, but at the same time the intensity of the emission is already nearly zero. The spectral line seen in emission between \ion{Ca}{ii} H and  H$\epsilon$ up to a height of nearly 2\,Mm should pertain to singly ionized iron \citep[cf.][]{engvold+halvorsen1973,lites1974,rutten+stencel1980,watanabe+steenbock1986,schmidt+fisher2013}.
%For the determination of kinetic temperatures and non-thermal velocities, the line shape should be roughly Gaussian, otherwise most of the assumptions underlying Eqs.~\ref{eq1} and \ref{eq2} should not be fulfilled. While for \ion{He}{i} at 1083\,nm a Gaussian shape can be attributed nearly down to the limb (taking the presence of the weaker blue component into account), \ion{Ca}{ii} H exhibits a central self-absorption core up to about 4\,Mm, a flat-topped profile from there to about 7\,Mm and a roughly Gaussian shape above this height.

\begin{figure}
%\hspace*{.5cm}\begin{minipage}{11.5cm}\resizebox{11.5cm}{!}{\includegraphics{gauss_rs_2d_zbar.ps}}\vspace*{.1cm}\\\resizebox{11.5cm}{!}{\includegraphics{gauss_rs_2d.ps}}\\$ $\\$ $\\\end{minipage}\hspace*{.5cm}\begin{minipage}{5.6cm}\end{minipage}
\centerline{\resizebox{12cm}{!}{\includegraphics{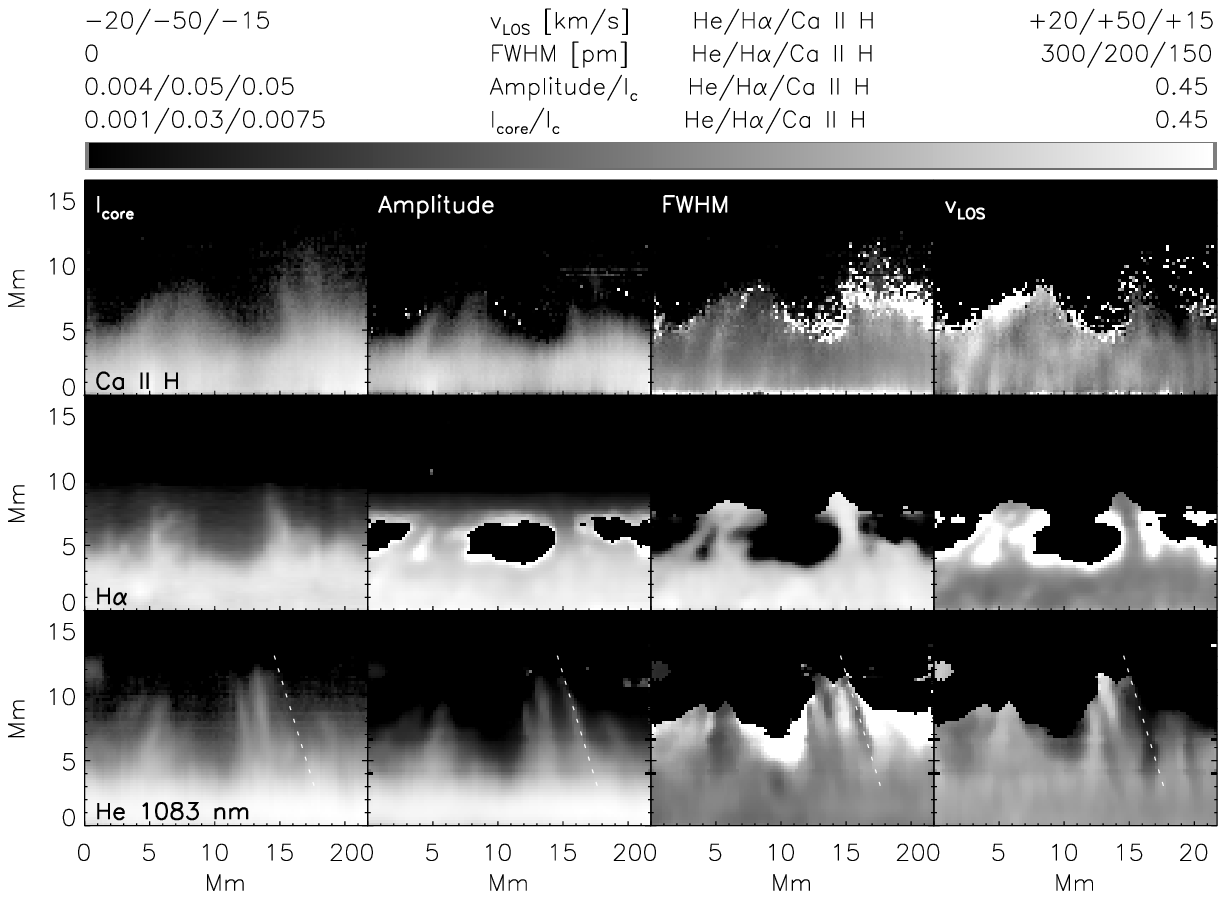}}}
\caption{Results of the Gaussian fit for observation No.~1. Left to right: line-core intensity, amplitude, FWHM and LOS velocity of the fitted Gaussian. {Top to bottom}: \ion{Ca}{ii} H, H$\alpha$, and \ion{He}{i} at 1083\,nm. The white inclined dotted line in the bottom row marks a region of reduced FWHM in \ion{He}{i} at 1083\,nm. The solar limb is at $y\,=\,$0\,Mm. The grey bar on top gives the display ranges for the respective parameters and lines.}\label{gaussres}

\end{figure}

Figure \ref{fwhm_allwl} shows the FWHM of the laterally averaged profiles for
\ion{He}{i} at 1083\,nm, \ion{Ca}{ii} H and H$\epsilon$. Apart from
H$\epsilon$, the general trend is a decrease in line width with height. For
H$\epsilon$, the line width increases on all three locations about 0.5\,Mm
in height before the emission has vanished completely. We note that this
apparent increase in line width in H$\epsilon$ happens at about 2--3\,Mm,
where all other lines indicate an usually monotic decrease in line width.

%, but we note that this only applies to the FWHM derived from the Gaussian fit. The simple method to determine the locations where the intensity has dropped to half of the maximal value indicates a monotonic decrease in line width in H$\epsilon$ for all significant spectra. \textbf{\textcolor{red}{([DF: ] The discussion in the previous sentence seems wrong to me. Are you talking about CA II H?)}}

Figure \ref{gaussres} shows all parameters retrieved from the Gaussian fit for
the \ion{He}{i}, \ion{Ca}{ii} H and H$\alpha$ spectra throughout the full FOV
of observation No.~1 as a cross-check of the behaviour of the line width in
individual spicules. Only the off-disc area of the FOV is shown. The line
width in \ion{Ca}{ii} H shows a faint lateral structuring (e.g., at $x\sim$\,4 to 8\,Mm), but no clear vertical variation. At best a weak trend for a reduction of the FWHM with height can be discerned. In H$\alpha$, the features at $x\sim\,5$ and 15\,Mm show an increase of the FWHM at the upper ends, but the intensity at these places is already low. Appendix \ref{signi} shows in more detail that and why the values derived from these profiles are spurious, whereas the Gaussian fit works acceptably well for profiles closer to the limb. In \ion{He}{i}, one out of the five distinct features that can be identified (two at $x\sim\,5$\,Mm, three at 15\,Mm) shows an increase of the FWHM at the upper tip similar to those seen in H$\alpha$. The FWHM increases up to the display threshold limit at the maximal height for which values are displayed in each column, but a comparison with the line-core intensity or the amplitude of the Gaussian reveals that these values are spurious because the intensity is nearly zero at these locations (cf.~between $x\sim$7 to 11\,Mm: large FWHM (white) at $y\sim7$ to 9\,Mm coincides with nearly zero amplitude of the Gaussian). The FWHM and the LOS velocities in \ion{He}{i} show similar patterns with more lateral fine-structure than the line-core image. The patterns of low/high FWHM or positive/negative LOS velocity are oriented similar to the spicules in the line-core images, with the same tilt relative to the limb and the same vertical extent. No clear relation between FWHM and the LOS velocities can, however, be derived instantly because all combinations of positive or negative LOS velocities with low or high FWHM can be found. The same holds for any possible relation of either FWHM or LOS velocities with high or low intensity.

Observations Nos.~2-4 are displayed in Appendix \ref{slit_add}. They yield the same result as obtained from observation No.~1: the line width decreases with increasing limb distance up to the point where the spectra show no significant emission anymore.

%{\em First column, top/bottom}: line width and amplitude of the Gaussian in the \ion{He}{i} spectra for comparison. {\em Second to seventh column}:  $T_{\rm kin}$ and $v_{\rm non-th}$ from the line pairs H$\alpha$\,--\,\ion{Ca}{ii} IR, H$\alpha$\,--\,\ion{Ca}{ii} H,  H$\alpha$\,--\,\ion{He}{i}, \ion{He}{i}\,--\,\ion{Ca}{ii} IR, \ion{He}{i}\,--\,\ion{Ca}{ii} H and the fit to all lines together.

\begin{figure}
%$ $\\$ $\\
%\hspace*{1cm}\resizebox{10.25cm}{!}{\includegraphics{macro_spic_fov_mmod.ps}}
\centerline{\resizebox{9cm}{!}{\includegraphics{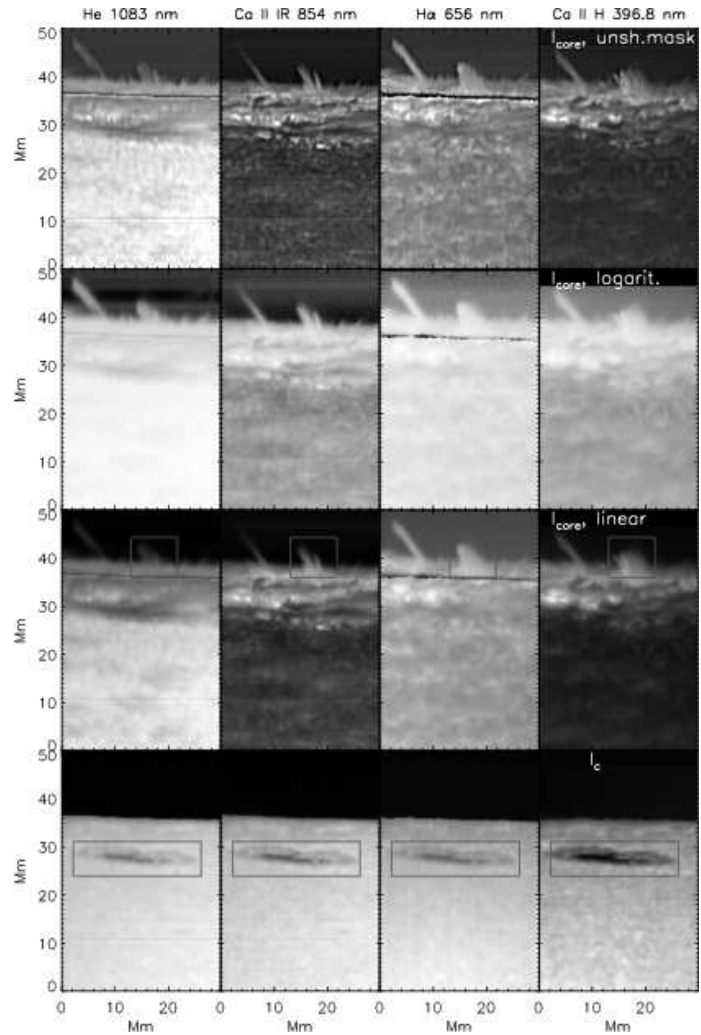}}}
\caption{Overview of the FOV of observation No.~5 taken on 12/04/2011. {Left to right}: \ion{He}{i} at 1083\,nm, \ion{Ca}{ii} IR at 854.2\,nm, H$\alpha$ and  \ion{Ca}{ii} H. {Bottom to top}: continuum intensity, line-core intensity in linear display, the same in logarithmic display and the same with unsharp masking. The red rectangles outline some features in the FOV that highlight the accuracy of the spatial alignment (bottom row) and the different appearance in the line core of different spectral lines (second row from the bottom).}\label{fov_good}
\end{figure}
\begin{figure}
%\begin{minipage}{11.8cm}\hspace*{1.25cm}\resizebox{10.cm}{!}{\includegraphics{macro_spic_gauss_zbar.ps}}$ $\\$ $\\\hspace*{.75cm}\resizebox{11.cm}{!}{\includegraphics{macro_spic_gauss.ps}}\\$ $\\\end{minipage}\hspace*{.3cm}

\centerline{\resizebox{12cm}{!}{\includegraphics{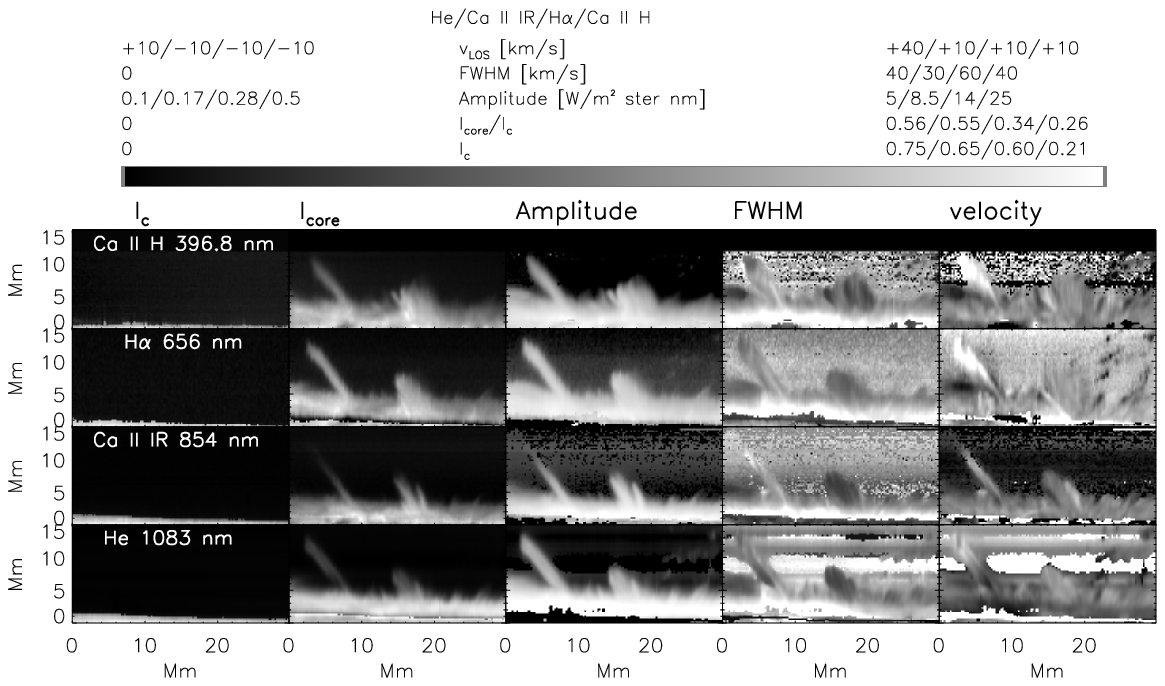}}}

\caption{Results of the Gaussian fit for observation No.~5. {Left to right}: continuum intensity, line-core intensity, amplitude, FWHM and LOS velocity. {Top to bottom}: \ion{Ca}{ii} H, H$\alpha$, \ion{Ca}{ii} IR at 854.2\,nm and \ion{He}{i} at 1083\,nm.} \label{gauss_res_good}
\end{figure}

The spatial resolution of all of the slit-spectrograph data shown up to
here was partially impeded by the limited performance of the AO system that
had to operate with a facula as the main feature for the correlation. A
comparison of the previous figures of the slit-spectrograph data with
Fig.~\ref{fov_good} at once shows the advantage of a better suited AO lock
point such as a sunspot. Apart from the top row of Fig.~\ref{fov_good}, the
images were constructed from the reduced and calibrated spectra with no
additional treatment apart from the stray-light correction. These data,
however, now correspond to an active region instead of the quiet Sun. The
bottom row of Fig.~\ref{fov_good} shows the continuum intensity in the full
FOV in \ion{He}{i} at 1083\,nm, \ion{Ca}{ii} IR at 854.2\,nm, H$\alpha$ and
\ion{Ca}{ii} H to facilitate a control of the spatial alignment. We used the
sunspot inside the red rectangle for that purpose. Because the differential
refraction was again rather large early in the morning when the corresponding
observations were taken, the \ion{Ca}{ii} H data are less well aligned
than all others, even if it is not obvious in the image. Features in the
\ion{Ca}{ii} H images are at the same place after the alignment, but the corresponding \ion{Ca}{ii} H spectra were actually taken about 70 (100) s earlier than those of H$\alpha$ (\ion{He}{i}). 

The line-core images of Fig.~\ref{fov_good} show several resolved spicules and
macrospicules of up to 20\,Mm extent at the limb. No relation of the off-limb
structure to the sunspot on the disc is directly obvious, e.g., by connecting
intensity brightenings or darkenings. The red rectangles in the second row
from the bottom outline a cluster of spicules. Their appearance changes
significantly between the different spectral lines, with both the \ion{Ca}{ii}
H and the IR line showing individual strands rather than the more uniform structure seen in H$\alpha$ or \ion{He}{i} at 1083\,nm. The time difference between the \ion{Ca}{ii} IR and H$\alpha$ (\ion{He}{i}) is zero (30\,s), so the different appearance in these three lines should not be caused by the temporal evolution. The largest macrospicule at $x\sim 5$\,Mm exhibits a substructure of a dark central core in the \ion{He}{i} line-core image, whereas the corresponding bright feature in \ion{Ca}{ii} IR at 854.2\,nm seems to correspond only to the central part of the macrospicule. 

Figure \ref{gauss_res_good} shows the results of applying the Gaussian fit to the spectra of observation No.~5. Only the off-limb region of the FOV is shown. We did not filter out the pixels without significant emission, where the results of the Gaussian fit are only spurious, as in all previous figures, because in all spectral lines the transition into noise can be clearly identified in one or all of the Gaussian parameters, best usually in the FWHM and LOS velocity. The FWHM in the cluster of spicules at $x\sim 15$ to 20\,Mm is similar for H$\alpha$, \ion{Ca}{ii} IR at 854.2\,nm and \ion{He}{i} at 1083\,nm, but shows a much stronger lateral structuring with an iterative change from low to high FWHM in \ion{Ca}{ii} H. In all lines, the FWHM decreases or remains at best constant with height above the limb for the cluster of spicules. In the largest macrospicule at $x\sim 5$\,Mm, the central axis shows an increased FWHM in all lines, and a reduction at the tip of the structure in all lines but \ion{Ca}{ii} IR at 854.2\,nm.
\begin{figure}
%\centerline{\resizebox{4.cm}{!}{\includegraphics{kin_temp_120411_zbar.ps}}}$ $\\\hspace*{.5cm}\resizebox{11.5cm}{!}{\includegraphics{kin_temp_120411.ps}}$ $\\$ $\\
\centerline{\resizebox{12cm}{!}{\includegraphics{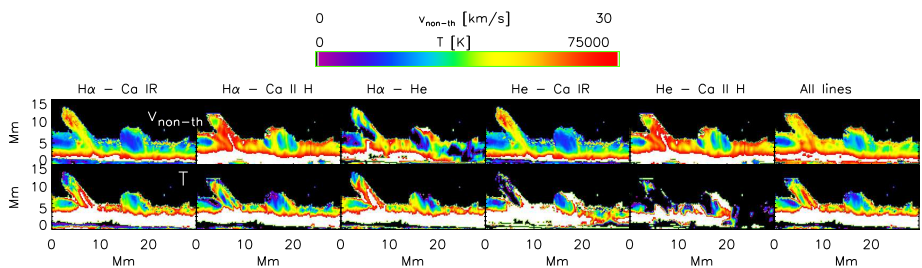}}}
\caption{Kinetic temperature $T_{\rm kin}$ (bottom row) and non-thermal velocity $v_{\rm non-th}$ (top row). {Left to right}:  $T_{\rm kin}$ and $v_{\rm non-th}$ from the line pairs H$\alpha$\,--\,\ion{Ca}{ii} IR, H$\alpha$\,--\,\ion{Ca}{ii} H,  H$\alpha$\,--\,\ion{He}{i}, \ion{He}{i}\,--\,\ion{Ca}{ii} IR, \ion{He}{i}\,--\,\ion{Ca}{ii} H and the fit to all lines together.}\label{last_gauss}
\end{figure}

We applied Eqs.~\ref{eq1} and \ref{eq2} to the roughly simultaneous,
co-spatial spectra of observation No.~5 to obtain an estimate of the kinetic
temperature and the non-thermal line width (Fig.~\ref{last_gauss}). We used
all line pairs of different chemical elements  and a least-square fit to all
lines simultaneously (rightmost column). The different line pairs give rise to
slightly different results, but some characteristics are common. The kinetic
temperature in the cluster of spicules at $x\sim 18$\,Mm ($<$\,30000\,K) is
lower than in the macrospicule at $x\sim 5$\,Mm (up to above 50000\,K). The
temperature generally reduces towards the upper end of both structures. A
similar reduction of kinetic temperature with height is to first order also
seen in the rest of the FOV, especially from the layers close to the limb ($y < 5$\,Mm) towards higher layers. The non-thermal line width increases slightly at the top of the macrospicule when the \ion{Ca}{ii} IR spectra are involved, whereas for all other line pairs and combinations it reduces. The non-thermal line width is about 5 to 10\,km\,s$^{-1}$ in the cluster of spicules and above 15\,km\,s$^{-1}$ in the isolated macrospicule. Across the FOV, the non-thermal line width decreases with height similar to the kinetic temperature, with a sharp drop at a height of about 5\,Mm. 
\begin{figure}
%\resizebox{11.6cm}{!}{\includegraphics{av_off_limb_spec_new.ps}}
\centerline{\resizebox{12cm}{!}{\includegraphics{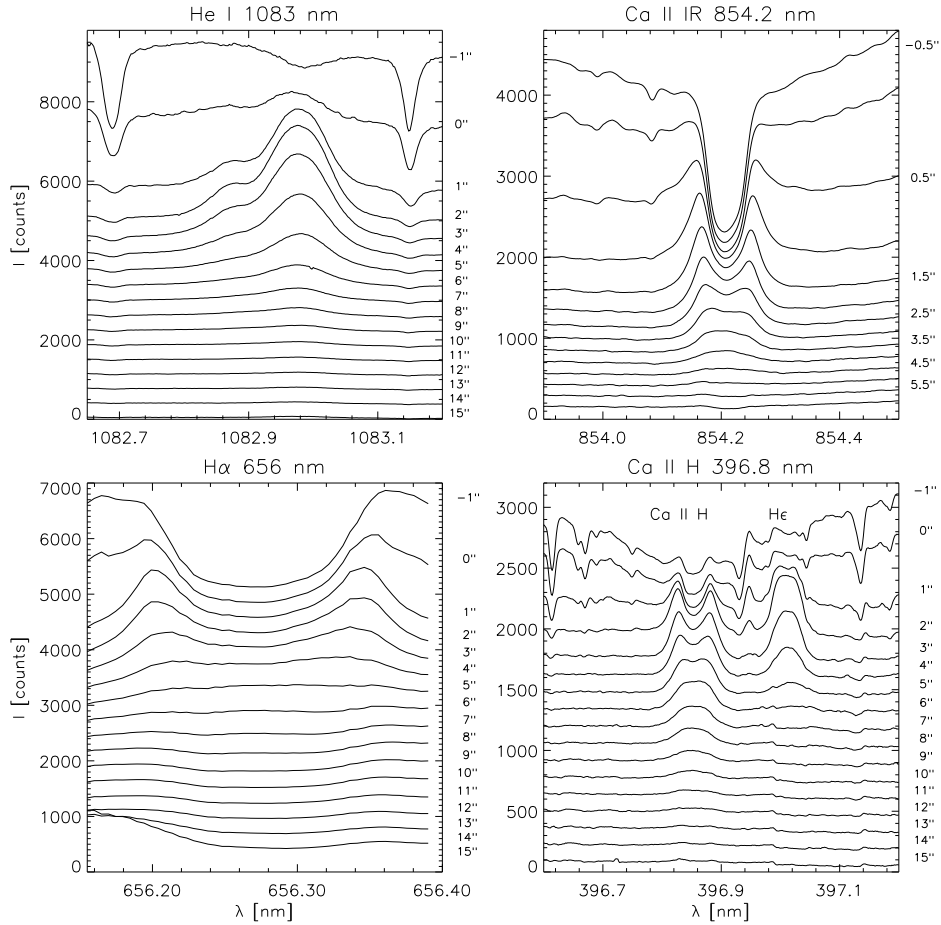}}}
\caption{Average off-limb spectra of observation No.1 (cf.~Fig.~\ref{all_wl}) of (clockwise, starting left top) \ion{He}{i} 1083\,nm, \ion{Ca}{ii} IR 854\,nm, \ion{Ca}{ii} H and
H$\alpha$. The location of the profiles relative to the limb is given
at the right-hand side of each panel. For \ion{Ca}{ii} IR at 854.2\,nm,
only profiles up to a height of 6$^{\prime\prime}$ are shown.}\label{av_prof_1} 
\end{figure}

\begin{figure}
%\resizebox{11.6cm}{!}{\includegraphics{av_prof_apr2011_new.ps}}\\
%\resizebox{11.6cm}{!}{\includegraphics{av_prof_apr2011_widthint_new.ps}}
\centerline{\resizebox{12cm}{!}{\includegraphics{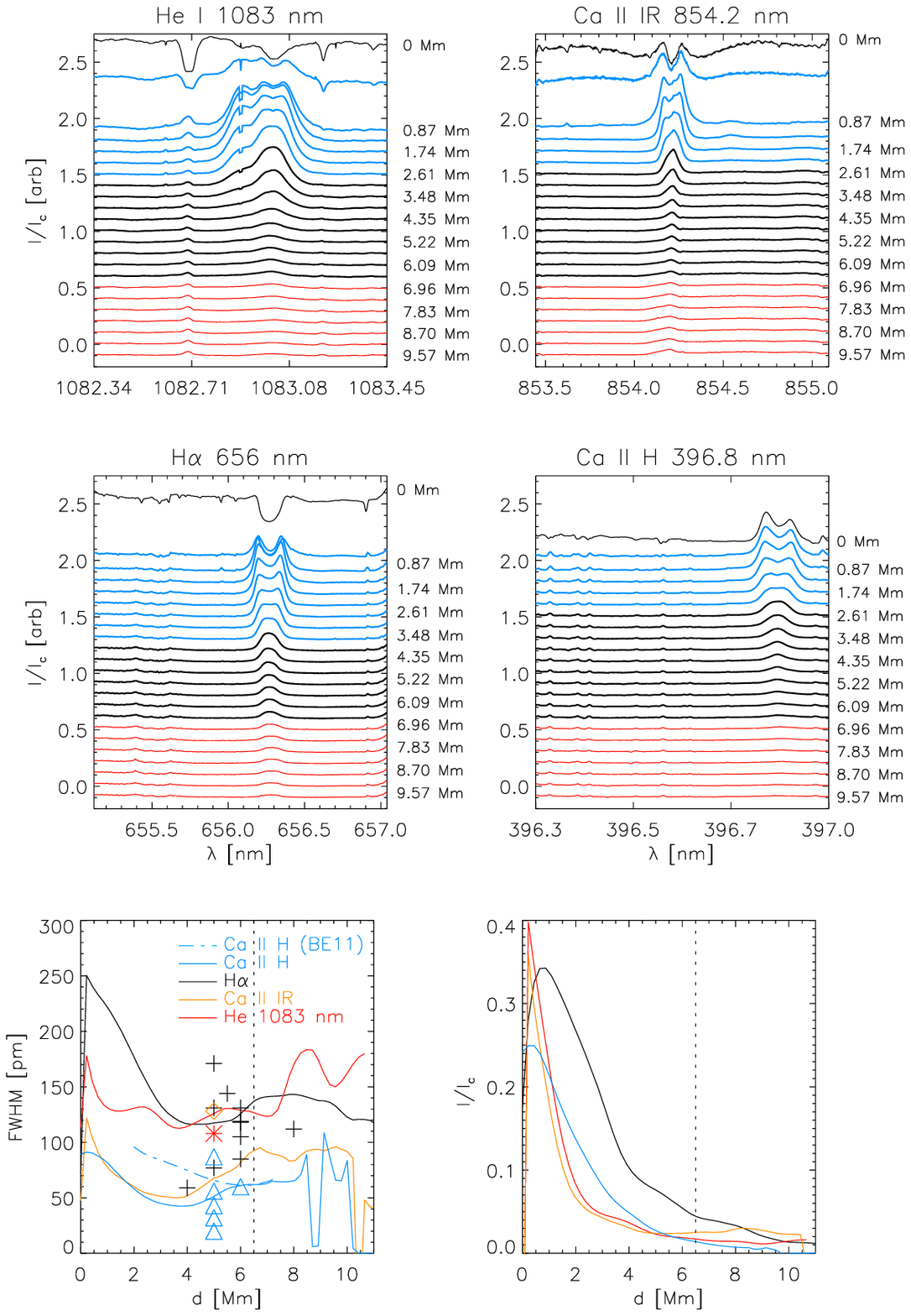}}}
\caption{{Top four panels}: average off-limb spectra of observation No.~5 (cf.~Fig.~\ref{fov_good}) of (clockwise, starting left top) \ion{He}{i} 1083\,nm, \ion{Ca}{ii} IR 854\,nm, \ion{Ca}{ii} H and H$\alpha$. The location of the profiles
relative to the limb is given at the right-hand side of each
panel. Profiles in blue show self-absorption or are flat-topped. Profiles in red are considered to represent spurious
intensities. {Bottom panels}: FWHM (left) and amplitude (right) of the
Gaussian for the average spectra. The red asterisk, orange diamond,
blue triangles and black crosses denote the FWHM values of \ion{He}{i}
1083\,nm, \ion{Ca}{ii} IR 854\,nm, \ion{Ca}{ii} H and H$\alpha$ given in
\citet{beckers68,beckers72} and \citet{alissandrakis1973}.}\label{av_2}
\end{figure}
\subsection{Average off-limb spectra}
Figures \ref{av_prof_1} and \ref{av_2} show the average off-limb spectra in
\ion{Ca}{ii} H, H$\alpha$, \ion{Ca}{ii} IR at 854.2\,nm and \ion{He}{i} at
1083\,nm for observations No.~1 and No.~5, respectively. The H$\alpha$ spectra
in Fig.~\ref{av_prof_1} were taken with TESOS and therefore cover only a
small wavelength range around the line core, whereas the \ion{Ca}{ii} H
spectra were recorded with the PCO inside of POLIS and also include the
H$\epsilon$ line. In Fig.~\ref{av_2}, the H$\alpha$ spectra were recorded with
a PCO and cover the full prefilter curve, whereas the H$\epsilon$ line was
not covered in the \ion{Ca}{ii} H spectra because of using the default POLIS Ca CCD camera. The displayed wavelength range was clipped to the line-core region in all plots.

All plots from both observations show the same general trend in line width
with increasing limb distance in all of the lines, i.e., the amplitude of the
emission reduces strongly, making some of them virtually disappear at a height
of 5\,Mm, and the line width reduces up to the point where the
amplitude is close to the significance limit, making the increase in line
width at the uppermost heights doubtful because of the presence of some
residual stray light (cf.~the photospheric lines in the wings of Ca and
H$\alpha$). The height range where the spectra are not significant is marked by red color in the top panels and by a dotted vertical line in the bottom panels of Fig.~\ref{av_2}, respectively. Profiles in blue show either self-absorption or are flat-topped which indicates line formation in an optically thick atmosphere (see Sect.~\ref{limitations} below).

For the data with the higher spatial resolution (observation No.~5 taken in 2011), we applied the Gaussian fit to the average profiles. The panels in the bottom row of Fig.~\ref{av_2} show the amplitude and the FWHM of the Gaussian for these average spectra. We extracted the line widths of the various lines from \citet{beckers68,beckers72} and \citet{alissandrakis1973} for comparison (small symbols in the lower-left panel of Fig.~\ref{av_2}). These values show a quite significant scatter that usually covers our results for the respective lines, where our data has the advantage of having a reliable limb distance and a complete coverage of the height range up to 10\,Mm above the limb. From the literature values, no clear trend in line width with height can be discerned. That the large scatter is actually a real problem caused by the use of different observations, recorded at different seeing conditions in  different regions on the Sun, can be seen
by comparing the line width of \ion{Ca}{ii} H with the one determined for the
same line in BE11 (dash-dotted line). Using the same instrument and wavelength
region, the line width retrieved in BE11 is larger than for observation No.~5
up to a height of about 6\,Mm. This could be caused by either the difference between active region or quiet Sun off-limb features, or the difference in seeing with its related spatial smearing of structures at different Doppler shifts that affects the line width.
\section{Summary\label{sec_summ}}
We have analyzed a set of multi-wavelength spectroscopic observations at the solar limb in some of the strongest chromospheric
spectral lines (\ion{Ca}{ii} H at 396.85\,nm, H$\epsilon$, H$\alpha$,
\ion{Ca}{ii} IR at 854.2\,nm and \ion{He}{i} at 1083\,nm). The setups used at
the German VTT to obtain simultaneous spectra in all the lines were
usually complex, making use of several instruments with additional cameras to cover more spectral lines. 

We analyzed the observed spectra on two main points, i.e., the spatial
structuring of the off-limb features and the line width of the spectral lines
as a function of their limb distance in individual spicules, across the FOV and in average profiles. A comparison of 2D spectroscopy in H$\alpha$ with slit-spectrograph
data in the other lines reveals that only H$\alpha$ and \ion{He}{i} at 1083\,nm show significant emission above a height of about 5 to
6\,Mm above the solar limb. Many of the structures seen in H$\alpha$ above this
height differ from the thin, elongated spicules seen
closer to the limb, i.e., they have lateral widths of up to a few Mm and
usually show  complex shapes. In observations at our spatial resolution of
about 1$^{\prime\prime}$ -- apart from the GFPI data with a spatial resolution
of about 0\farcs3 -- , limb spicules
merge into a dense forest without individual structure in line-core images up
to a height of about 6\,Mm, whereas in the Doppler velocity some spicules can
still be identified at lower heights. Spicules extending above a height of
6\,Mm can be seen as isolated, individual features. The lateral structuring is
more pronounced than the vertical structuring, with some features maintaining
their small lateral width of 1\,Mm or less over a length of several
Megameters. Large-scale structures in H$\alpha$ exhibit brightenings that
propagate upwards, mostly in the form of roundish blobs. These features
usually appear only in connection with large-scale, complex-shaped structures that extend beyond the typical height range of spicules.

The line width in individual spicules reduces with the height above the limb
in most cases. Some lines maintain a roughly constant line width over a height
range of a few Mm. We find an increase in the line width at the uppermost tips
of spicules, but its significance is doubtful because the corresponding
amplitude of the emission is at the noise level of the spectra. Our
results for the line width in spicules are covered by values published in
previous literature, but both the literature values and our own measurements
at two different times show a large scatter.

The derivation of kinetic temperatures and non-thermal velocities from the set
of the best simultaneous spectra yields temperatures of about 10000 to
70000\,K for spicules and macrospicules, respectively. The non-thermal velocities are between 10 and 30\,km\,s$^{-1}$. These values refer, however, to off-limb
structures seen near or in an active region. For our quiet Sun data, we have no equivalent set of
co-spatial and simultaneous spectra because of a significant temporal shift
between different wavelengths caused by the combination of differential
refraction and sequential scanning. The main reason for this drawback is that
the POLIS slit could not be rotated to orient it perpendicular to the horizon, while additionally the location of POLIS in the observing room of the VTT caused the largest differential refraction effects early in the morning \citep[cf.~Appendix A of][]{beck+etal2007}.
\section{Discussion\label{sec_disc}}
For the observation of individual spicules, a high spatial resolution is
required. Figures \ref{all_wl} and \ref{fov_good} demonstrate that thanks to
adaptive optics, ground-based spectrograph data with an integration time
of a few seconds can achieve this.  The pseudo-scan map of H$\alpha$ in
Fig.~\ref{all_wl} additionally proves that the slit-spectrograph data
correspond to a correct sampling of the temporal evolution seen in the
simultaneous 2D spectroscopic data, with the limitation for their
interpretation that they only cut out a single position from a temporally
fast evolving 2D pattern. To preserve the spectral information, i.e., the
shape of the emission and any eventual Doppler shifts, it is necessary to
integrate over a few seconds because of the low light level of down to a few
percent of the disc-centre intensity. Broad-band filter observations do not
maintain the spectral information, but allow one to use shorter exposure or
integration times. They provide the option to study the temporal evolution in
detail \citep[e.g.,][]{pereira+etal2012} but do not allow one to extract physical properties of the solar atmosphere from the data themselves, making the review of \citet{beckers68}, which summarizes results of data taken before, still the reference for the physical properties of spicules.

The extraction of physical properties also depends to some extent on the
availability of multiple spectral lines. The distinction between thermal and
non-thermal line width, or any more detailed modeling and analysis requires
more information than a single spectral lines usually provides, with the
exception of the \ion{He}{i} line at 1083\,nm that in itself has several components \citep{sanchezandrade+etal2007,centeno+etal2010,martinezgonzalez+etal2012}.
Such multi-wavelength capabilities are therefore important to preserve and
provide in future solar-telescope projects such as the DKIST
\citep{rimmele+etal2010} or EST \citep{collados+etal2010}. It seems that  in
the last few decades similar multi-wavelength data have mainly been recorded for prominences (far) off the solar limb \citep[e.g.,][]{stellmacher+wiehr1981,bendlin+etal1988,stellmacher+etal2003}, but not for structures near and at the limb \citep[apart from][]{socasnavarro+elmore2005}. 

Stray light off the limb is an issue that needs to be dealt with in spicule
observations. In the case that no direct measurements of the point-spread
function are available \citep[cf.][]{beck+etal2011,loefdahl+scharmer2012}, a
wavelength region in the continuum, where  off the limb no solar emission is
expected, can be used for the stray-light correction. For off-limb observations in chromospheric lines with 2D spectrometers such as TESOS, IBIS, CRISP or the GFPI, such a continuum wavelength point can be simply added to the spectral observation sequence to precisely determine the limb location and to correct for the off-limb stray light. 

We found a limited extent of about 5 to 6\,Mm above the limb for spicules, if
they are defined as elongated, thin intensity streaks. Only a few isolated
spicules can reach up to a height of 10\,Mm. Most other features seen
above 6\,Mm are complex, large-scale structures with a significant lateral
extent of a few Mm. There are no clear indications for any increase in line
width at the tip of spicules, neither in individual spicules nor on
average in the spectral lines used in our study (\ion{Ca}{ii} H, H$\epsilon$,  H$\alpha$, \ion{Ca}{ii} IR at 854.2\,nm, \ion{He}{i} at 1083\,nm). The same general decrease of line width in average spectra is also seen in \ion{Mg}{ii} spectra obtained with IRIS \citep[][their Fig.~4]{pereira+etal2014}. Structures above a height of 6\,Mm do not exhibit a significantly larger line width than is seen below that height. 

In our analysis of the line width, we did not try to distinguish between
  different types of spicular structures. This would have been possible only in
the H$\alpha$ time-series from TESOS or the GFPI. One additional reason is that
the distinction between the two types is still under debate
\citep{zhang+etal2012,skogsrud+etal2015}. If the type II spicules are dominant
in quiet Sun regions, they then also should have an impact on average
profiles or show up prominently across the field of view (e.g., Fig.~\ref{spic_evol_gfpi}).

\subsection{Limitations of the current analysis \label{limitations}}
Our current analysis has a few limitations for technical reasons and because of the intrinsic properties of the spectral lines observed. The low light level off the limb -- actually going down to zero -- introduces significant noise in the spectra. In addition, the stray light off the limb cannot be fully corrected for. The automatic analysis using a single Gaussian partly fails at emission amplitudes below 5\,\% of $I_c$ (Appendix \ref{signi}). However, the most stringent limitation is that all of the lines form in an optically thick regime in the lower part of the atmosphere. A clear indication for that is the appearance of self-absorption or flat-topped profiles (Fig.~\ref{av_2}) because in the optically thin regime a Gaussian or Voigt profile shape should prevail. For the strong chromospheric lines (all but H$_\epsilon$), the profiles indicate an optically thick regime up to a height of about 3\,Mm above the limb. Our estimates of the line width, regardless whether by the direct method or the Gaussian fit, overestimate the line width in the height range of optically thick line formation because the true maximal emission amplitude cannot be determined. We manually fitted single Gaussians to a few profiles with clear self-absorption by forcing the amplitude of the Gaussian to be larger than the observed peak emission values and adjusting the width by hand. This yielded a line width smaller by 20\,\% (8\,\%) at a height of 2.6\,Mm (3.6\,Mm) above the limb than for the automated single-Gaussian fit . The general decrease in line width with height above the limb was maintained, although with a smaller slope. The best way to avoid the influence of the line formation on the derivation of line width -- which is interpreted as reflecting temperature -- would be a full non-local thermodynamic equilibrium modeling of spicules at different temperatures and subsequent profile synthesis as done in \citet[][e.g., their Fig.~3]{judge+carlsson2010}, but for all lines of the current observations. The interpretation of line width as temperature measure or the derivation of a kinetic temperature from the width of lines from different chemical elements come with some uncertainty \citep{stellmacher+wiehr2015}.

%We thus do not see indications that spicules are {\em gradually} heated to transition region or coronal temperatures nor that they usually continue beyond a height of 6\,Mm. 
%The structures that are seen above the limb in the spectral lines used in our study (\ion{Ca}{ii} H, H$\epsilon$,  H$\alpha$, \ion{Ca}{ii} IR at 854.2\,nm, \ion{He}{i} at 1083\,nm) give the impression of being caused by cool chromospheric material such as the one in prominences. 
\begin{figure}
%\resizebox{12cm}{!}{\includegraphics{cair_rvdv09.ps}}\\
%\resizebox{12cm}{!}{\includegraphics{halpha_rvdv09.ps}}\\$ $\\
\centerline{\resizebox{14cm}{!}{\includegraphics{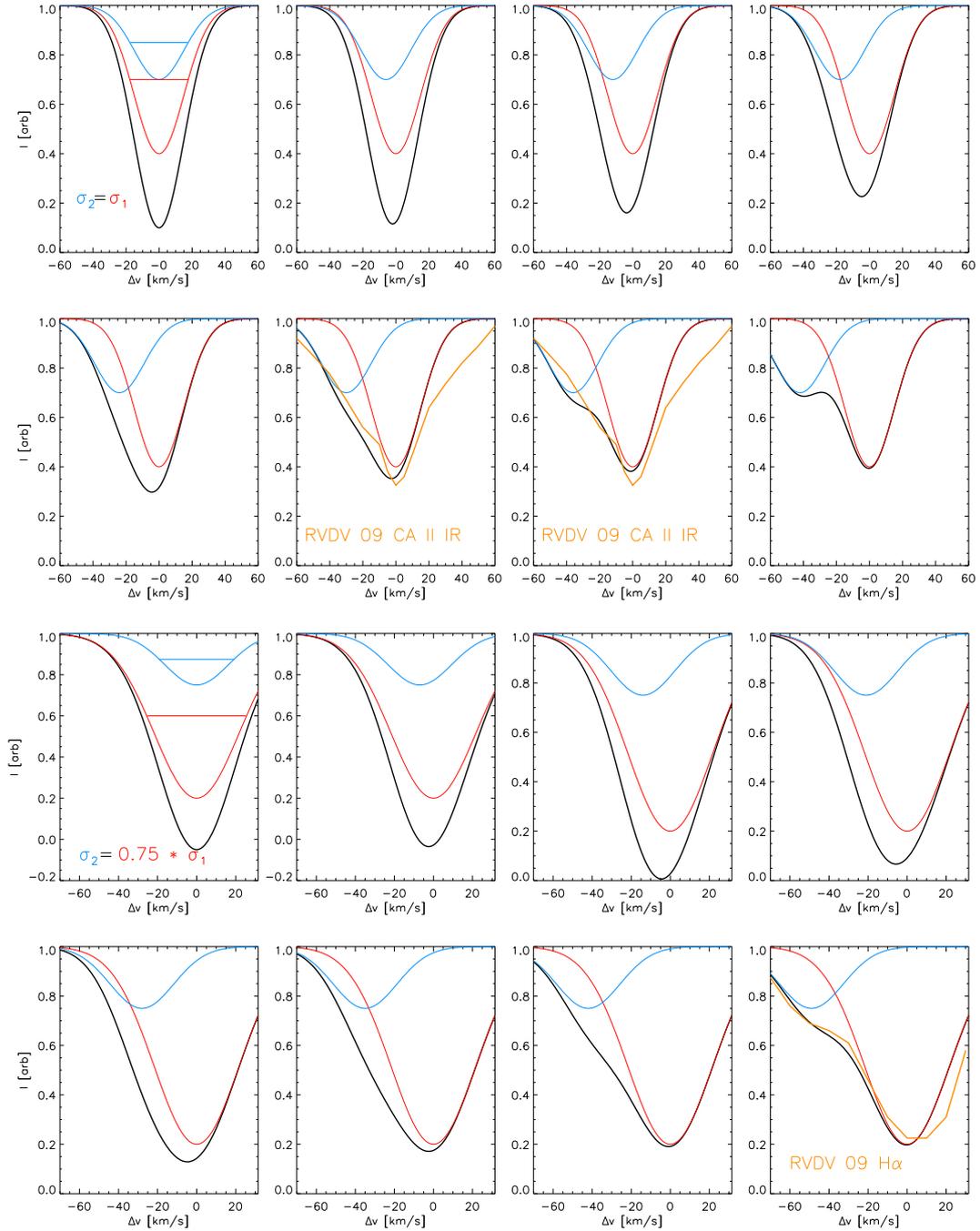}}}
\caption{Reproduction of \ion{Ca}{ii} IR (top two rows) and H$\alpha$ (bottom two rows) profiles observed in RBEs on-disc. Two Gaussians are overlayed in each panel. The red line corresponds to an un-shifted dominant component with a width $\sigma_1$, the blue line to a Doppler-shifted component with a width $\sigma_2$. The Doppler shift increases from left to right in each row. The orange line corresponds to the RBE profiles of \ion{Ca}{ii} IR and H$\alpha$ in Fig.~11 of \citet{rouppe_leen_etal09}. In the case of  \ion{Ca}{ii} IR, the observed profile can be reproduced to first order with Gaussians of the same width, while for H$\alpha$ a very good reproduction is achieved even if the width of the Doppler-shifted component is only 75\,\% of that of the component at rest.} \label{fig_kill}
\end{figure}

\subsection{Comparison to Rapid Blue Events}
 How can our finding of a decrease in line width with height above the limb be
 reconciled with that of \citet{rouppe_leen_etal09}
 that rapid blue events (RBEs), the counterpart of spicules on the disc, show
 an increase in line width at their tips ? For that, one has to take three
 points into account. Firstly, they found a positive correlation between blue
 shift and line width (e.g., their Fig.~11); secondly, one has to consider
 their equation to determine the line width; and finally, one has to consider
 the shape of the profiles they find in RBEs (e.g., their Figs.~5, 9 and
 11). Their RBE profiles show clear indications of two different components: a
 dominating component with a small -- or none at all -- Doppler shift and a
 strongly blue-shifted satellite of lower line depth \citep[see
 also][]{peter2001,tian+etal2011}. Their equation to determine the line width
 does not take the existence of the two components fully into account and
 corresponds more to a measure of line asymmetry than line width. Considering the line shape of the RBE profiles, the positive
 correlation between the velocity and the width follows directly: the larger
 the Doppler shift of the blue-shifted component, the larger the line
 asymmetry. Figure \ref{fig_kill} shows that asymmetric RBE profiles as given
 in \citet{rouppe_leen_etal09} can be generated without any increase in the
 line width in the Doppler-shifted component (compare also with the
 \ion{He}{i} spectra in Figs.~\ref{indi_allwl}, \ref{av_prof_1} and \ref{av_2} of this paper that
 {\em do} consist of two components). In the case of their RBE H$\alpha$
 profile, one can achieve a very good reproduction of the observed profile
 even if the width of the Doppler-shifted component is 25\,\% {\em smaller}
 than that of the un-shifted component. We thus conclude that if such
 asymmetric profiles are typical for RBEs, they must not indicate an increased
 line width in the Doppler-shifted component, and hence also do not provide
 direct evidence for heating in RBEs.

\subsection{Spicule disappearance}
The rather sharp boundary in height up to which spicules reach, the change
of the line shape of the spectra with height and the disappearance \textit{in
  situ} can be explained by  different effects. On the one hand, the emission
in the respective line cores can disappear because there is not enough
material left to generate a sufficient opacity, and hence the amplitude of the
emission reduces smoothly to zero (cf.~Fig.~\ref{av_2}). Because the density
stratification in the quiet solar atmosphere can be assumed to be spatially
rather homogeneous on large spatial or temporal scales, this would explain the
sharp and well-defined upper limit in height above which almost no spicules
can be seen. A second possibility is that the material would get heated
strongly and rapidly such that the necessary absorbers (or emitters) are
lacking because the elements that cause the spectral lines get ionized to higher ionization states. This
scenario has been invoked to explain spicules that fade away \textit{in situ}
\citep{depontieu_macintosh_etal07,sterling+etal2010} and implies that spicular
material could act as energy reservoir for the corona
\citep{depontieu+etal2009,depontieu+etal2011}. The fact that counterparts to
spicules are not always found in coronal lines
\citep[e.g.,][]{madjarska+etal2011} could be by chance \citep[but see
also][]{klimchuk2012}. Finally, a third option would be that spicules consist
of material that is shot up into the chromosphere to some height and that is
hotter than the mass in the surroundings. The excess energy  would then be
lost by radiative cooling in the presumably optically thin environment, while
the mass generally would follow a parabolic path of ascent and descent. On its path, the mass pertubation would temporally increase the density, hence the opacity and the emitted radiation would increase. That would make the travelling mass visible in the same way as prominences or coronal rain \citep{antolin+etal2012,oliver+etal2016} show up above the solar limb. It would lead to a sequential appearance and disappearance of spicules in low-forming (\ion{Ca}{ii}) and high-forming lines (\ion{Mg}{ii}, transition region) as described in \citet{skogsrud+etal2015} in terms of a passage of the material without requiring heating.

Can our multi-line spectra be used to distinguish between these possibilities\,?
There is one argument in favor of the reduction of density as reason for the
disappearance of spicules above a height of 6\,Mm, opposite to a rapid and
strong heating to transition-region or coronal temperatures. The emission in both the \ion{Ca}{ii} IR line at 854.2\,nm and the H$\epsilon$ line disappears in a completely analogous way as in \ion{Ca}{ii} H, H$\alpha$ or \ion{He}{i} at 1083\,nm by smoothly reducing to zero with height. The former lines, however, disappear already at heights below 4\,Mm (cf.~Fig.~\ref{av_prof_1}). It is then highly unlikely that a rapid and strong heating at a height of 4\,Mm would only affect these two lines without any impact on the other lines from the same chemical elements, or without changing the line width of \ion{He}{i} at 1083\,nm at this height. We thus suggest that the sharp upper boundary for the appearance of spicules, if defined as elongated, thin structures, is only the consequence of the lack of emitters caused by the reduction of density, and does not indicate a heating to transition-region temperatures. 
\section{Conclusions \label{sec_concl}}
In an analysis of multi-wavelength observations (\ion{Ca}{ii} H at 396.85\,nm,
H$\epsilon$, H$\alpha$, \ion{Ca}{ii} IR at 854.2\,nm and \ion{He}{i} at
1083\,nm) of solar spicules at the limb, we find that spicules corresponding
to the recently introduced type I or type II classes  extend only up to a height of about 5 to 6\,Mm. Structures above this height are mainly seen only in H$\alpha$, and differ by a significantly larger lateral width of up to a few Mm. All of the spectral lines show a decrease in the line width with height both in
individual spicules, across the FOV and in average profiles. A slight reversal
of the trend at the uppermost tips happens in spectra with very low
intensities and was found to be spurious, being caused by the limitations of the data and their analysis. A derivation of the kinetic temperature and non-thermal velocities yields a decrease in both quantities with increasing limb distance. We thus find no indications that the spicules in our data are gradually or rapidly heated to transition-region or coronal temperatures or that they extend to coronal heights.
\begin{acknowledgements}
The VTT is operated by the Kiepenheuer-Institut f\"ur Sonnenphysik (KIS; Freiburg, Germany) at the Spanish Observatorio del Teide of the Instituto de Astrof\'{\i}sica de Canarias (IAC; La Laguna, Tenerife, Spain). POLIS was a joint development of the High Altitude Observatory (HAO; Boulder, USA) and the KIS. Hinode is a Japanese mission developed and launched by ISAS/JAXA, with NAOJ as domestic partner and NASA and STFC (UK) as international partners. It is operated by these agencies in co-operation with ESA and NSC (Norway). SOHO is a project of international cooperation between ESA and NASA. HMI data are courtesy of NASA/SDO and the AIA, EVE, and HMI science teams. R.R.~acknowledges financial support by the Deutsche Forschungsgemeinschaft under grant RE 3282/1-1. D.~F. acknowledges financial support by the Spanish Ministries of Research and Innovation and of Economy through projects AYA2011-24808 and CSD2007-00050.
\end{acknowledgements}
\bibliographystyle{spr-mp-sola}
\bibliography{references_luis_mod1}
\begin{appendix}

\begin{figure*}
\centerline{\resizebox{12cm}{!}{\includegraphics{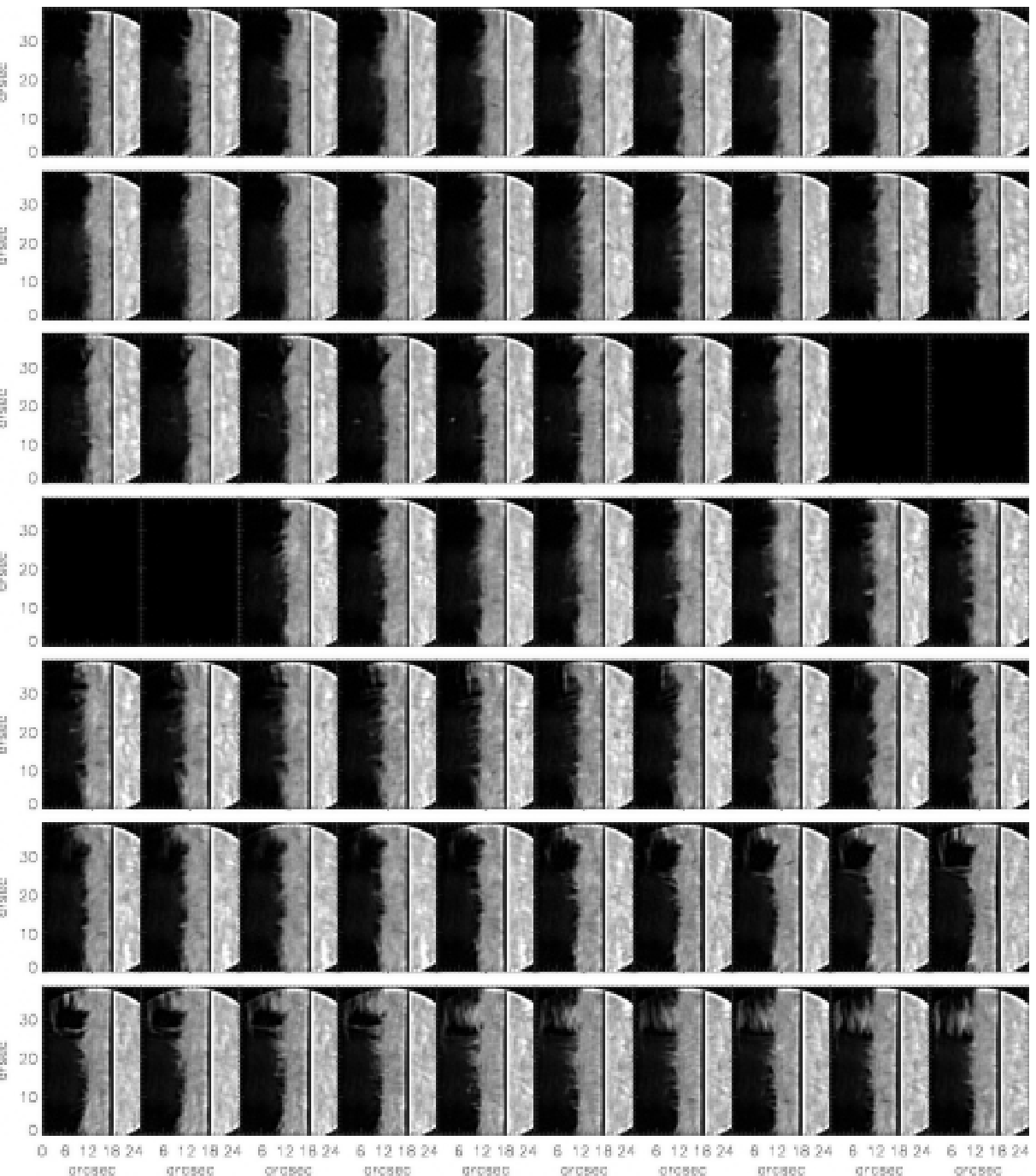}}}
\caption{First half of the time series of H$\alpha$ line-core images corresponding to observation No.~2 in Table \ref{tab_obs}. Only the image with unsharp masking is displayed. Time increases from left to right and top to bottom. The cadence between subsequent images is about 20\,s.} \label{ha_1}
\end{figure*}
\begin{figure*}
\centerline{\resizebox{12cm}{!}{\includegraphics{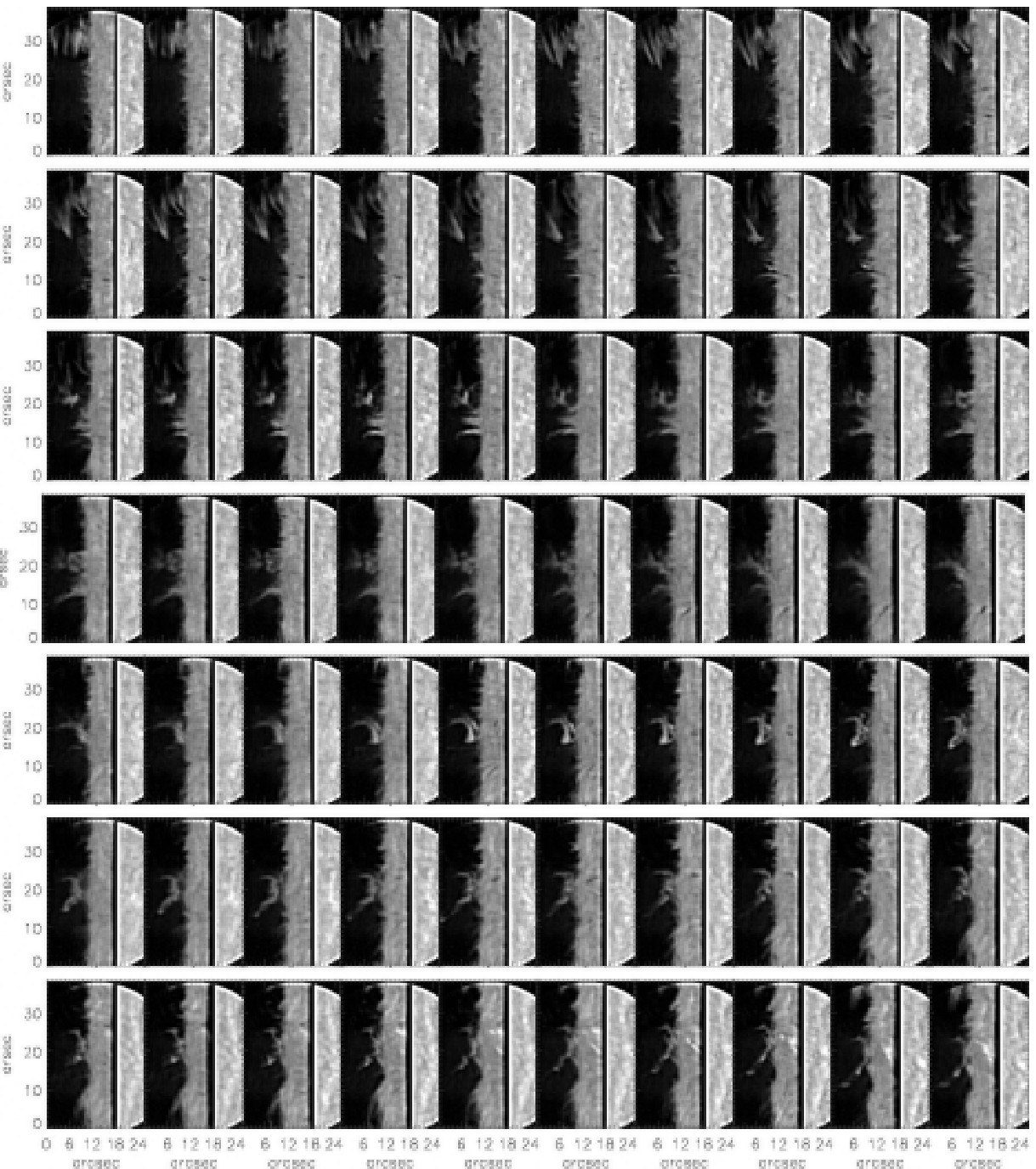}}}
\caption{Second half of the time series of H$\alpha$ line-core images corresponding to observation No.~2 in Table \ref{tab_obs}. Only the image with unsharp masking is displayed. Time increases from left to right and top to bottom. The cadence between subsequent images is about 20\,s.}\label{ha_2}
\end{figure*}

\begin{figure*}

\centerline{\resizebox{12cm}{!}{\includegraphics{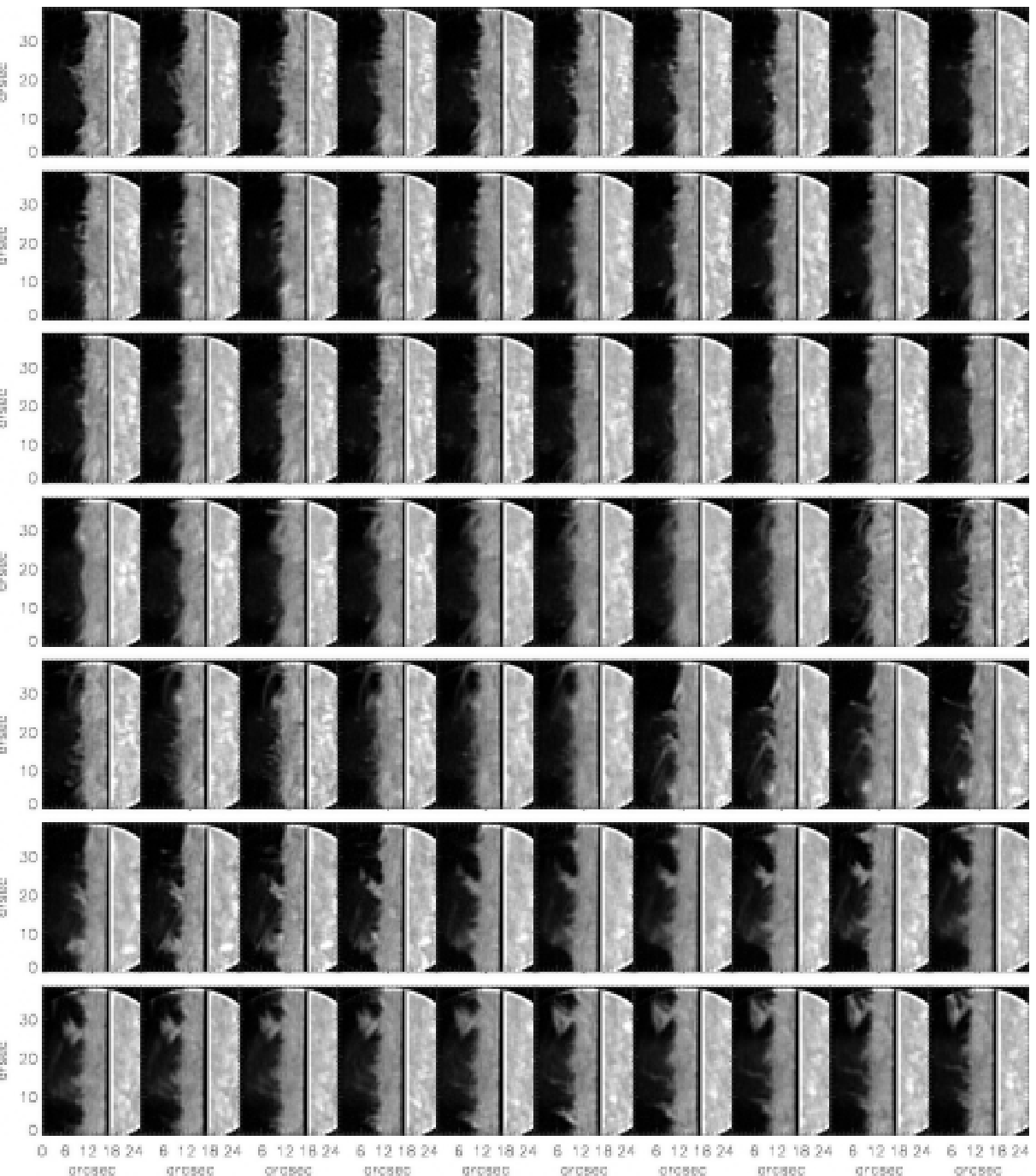}}}
\caption{First part of the time series of H$\alpha$ line-core images corresponding to observation No.~3 in Table \ref{tab_obs}. Only the image with unsharp masking is displayed. Time increases from left to right and top to bottom. The cadence between subsequent images is about 20\,s.}\label{ha_3}
\end{figure*}
\section{2D spectroscopy in H$\alpha$\label{app_ha}}
\begin{figure*}
\centerline{\resizebox{12cm}{!}{\includegraphics{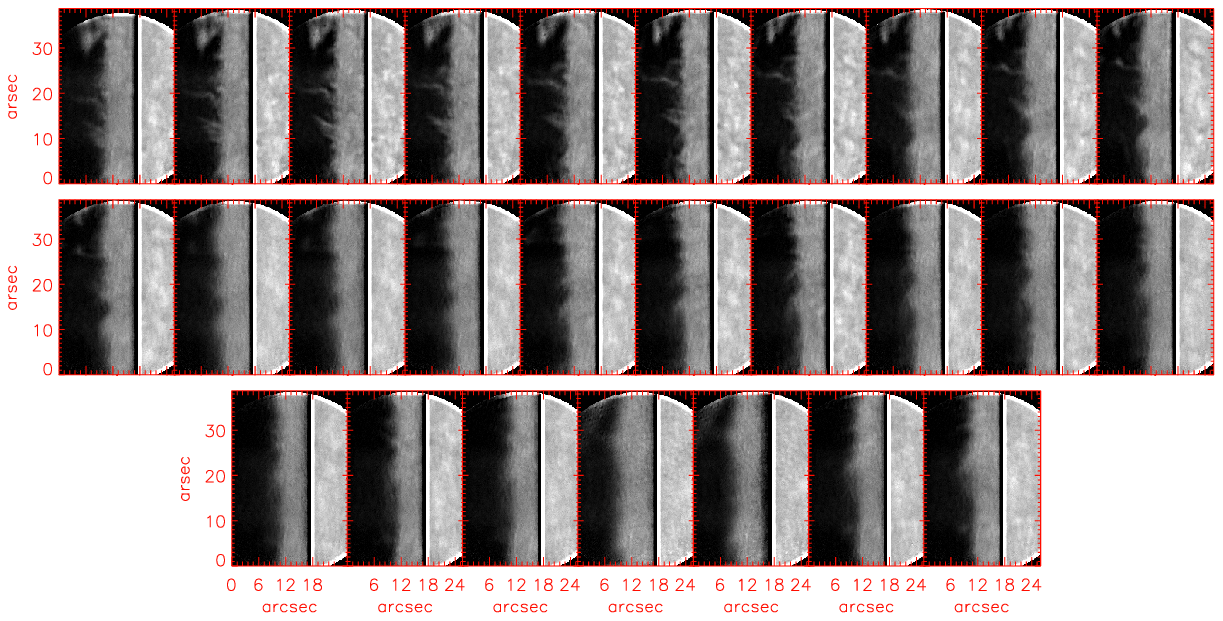}}}
\caption{Second part of the time series of H$\alpha$ line-core images corresponding to observation No.~3 in Table \ref{tab_obs}. Only the image with unsharp masking is displayed. Time increases from left to right and top to bottom. The cadence between subsequent images is about 20\,s.}\label{ha_4}
\end{figure*}

Figures \ref{ha_1} to \ref{ha_4} show the H$\alpha$ line-core images recorded
with TESOS during observations Nos.~2 and 3 in setup 1. Only the unsharp-masked images are shown. Both series are dominated by the appearance of large-scale
structures of a few Mm lateral width at heights above 6\,Mm. The first example (Figs.~\ref{ha_1} and \ref{ha_2}) shows a complex evolution, where some material seems to be falling down towards the limb, while at the same time several brightenings
start to protrude next to the location where the falling mass reaches it lowest
height. These two features, the falling mass and the extending brightenings,
seem to merge together afterwards and eject a blob of material towards the
corona. The ejected material rises upwards until it leaves the FOV of the
instrument. In the second example (Figs.~\ref{ha_3} and \ref{ha_4}), a
large-scale triangular structure with a large diameter above and a small
diameter near the limb forms from a tree-like structure with two branches. The
scenery is relatively quiet all throughout the FOV before, whereas with the
appearance of the large-scale structure, several individual features
appear above a height of 6\,Mm, affecting basically the full FOV of
more than 20\,Mm lateral extent at the same time. The large-scale structure is then unfortunately moved out of the FOV by the sequential scanning of the solar
image by the spectrograph instruments, so the return to a relative quiet scene
at the end could also be caused by this reason rather than a subsiding of the
activity. The last images also show the degradation of the seeing during the observations.

%\clearpage
\section{Limitations of Gaussian fits and significance of results \label{signi}}
\subsection{Single vs.~double-Gaussian fit}
\begin{figure}
%\hspace*{1cm}\resizebox{10cm}{!}{\includegraphics{indivspec3.ps}}$ $\\$ $\\
\centerline{\resizebox{9cm}{!}{\includegraphics{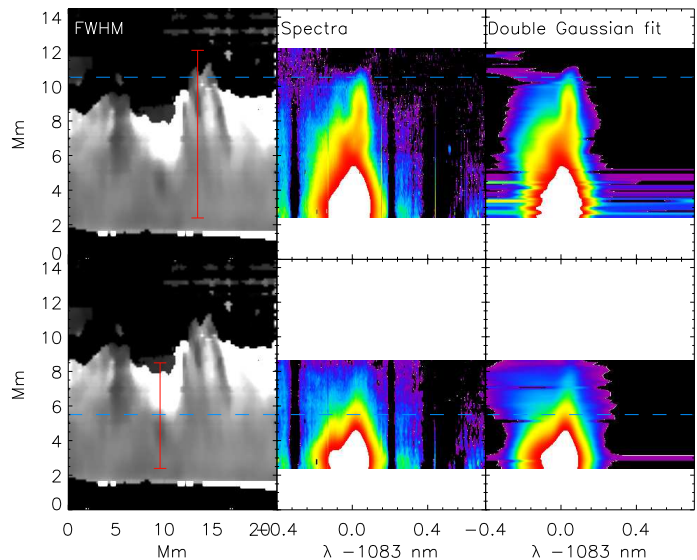}}}
\caption{Quality of double-Gaussian fits. Left: FWHM from single-Gaussian fit. The red vertical bars mark the spatial location of the spectra shown in the other columns. The horizontal blue dashed lines denote the maximum extent where the fits are deemed significant and reliable. Right two columns: spectra and double-Gaussian fits along the red bars in the left column. The values are displayed on a logarithmic scale, clipped and in false color to highlight the shape at the lowest intensities.}\label{lim1}
\end{figure}

The generally low light level beyond the limb, the decrease in the emission and the residuals of the stray-light correction hamper any reliable analysis. For all observations and all spectral lines, a single-Gaussian fit to all off-limb spectra was performed. All spectra with an amplitude of the Gaussian below a manually set threshold were rejected in order to remove profiles that contained only noise. To ensure that all significant profiles were kept, this threshold was set slightly lower than what would in principle be normal. For instance, for the He 1083\,nm spectra of observation No.~1 the threshold level was chosen as 0.4\,\% of $I_c$, leading to the uniform black area in FWHM in Fig.~\ref{gaussres}, third column, bottom panel. With this initial threshold, regions with a strongly increased FWHM in the results of the single-Gaussian fit remained as being significant (Fig.~\ref{lim1}). In the following, we investigate the corresponding profiles in more detail to show that the FWHM values derived from these profiles are spurious.
%This initial threshold left, however, regions with a strongly increased FWHM in the single-Gaussian results as significant (Fig.~\ref{lim1}). We took a closer look at the corresponding profiles to show why these values are spurious. 
\begin{figure}
%\resizebox{12.cm}{!}{\includegraphics{indivspec1.ps}}
\resizebox{12.cm}{!}{\includegraphics{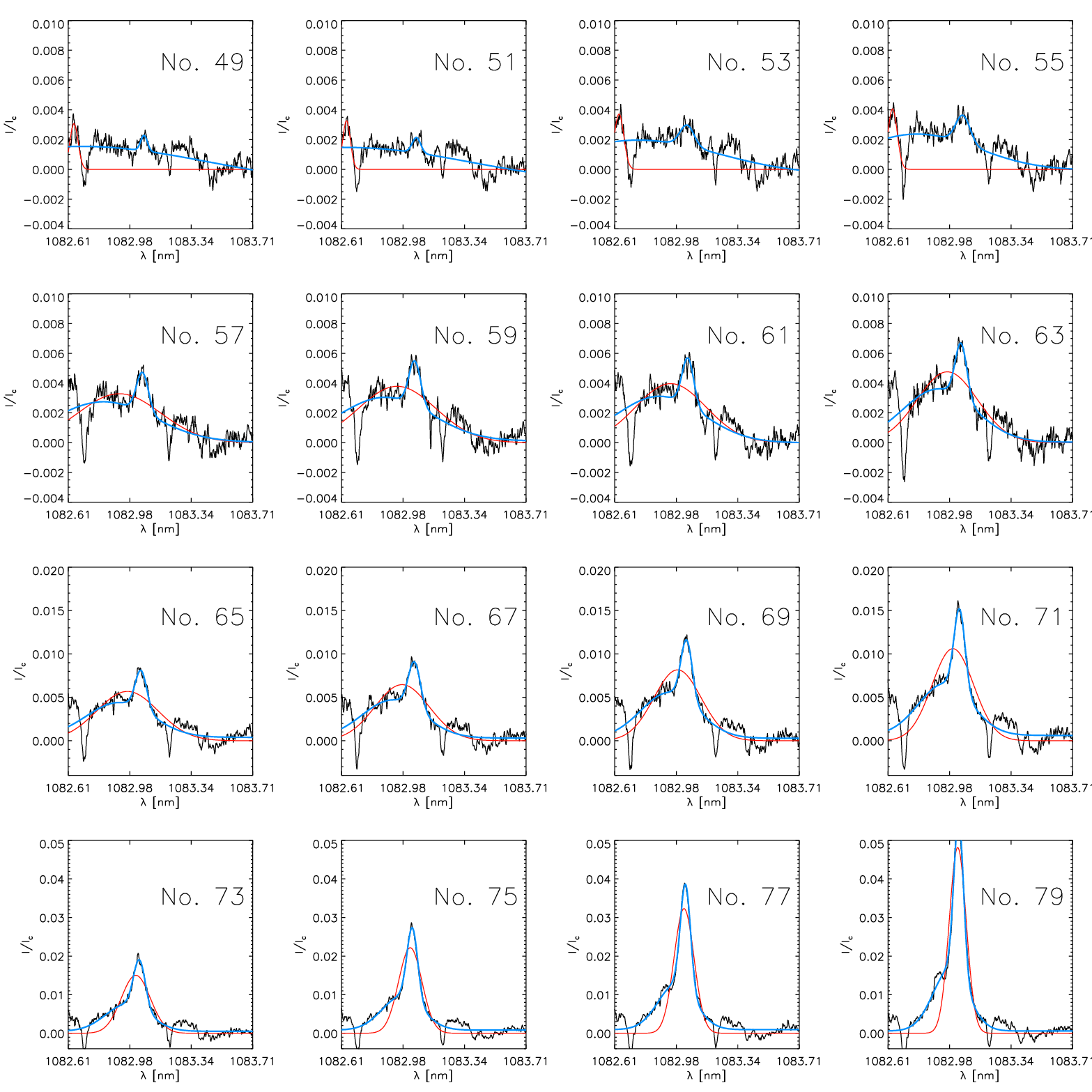}}
\caption{Gaussian fits to individual spectra. The observed spectra along the red bar in the lower row of Fig.~\ref{lim1} are shown with black lines. Red/blue lines show the results of the single/double-Gaussian fits. The two components of the double-Gaussian fit are shown separately in Fig.~\ref{lim3}. The single-Gaussian fit works up to about pixel No~73. Decreasing numbers indicate increasing limb distance.}\label{lim2}
\end{figure}

Figure \ref{lim1} shows the FWHM of He 1083\,nm obtained from the single-Gaussian fit for observation No.~1. We selected two vertical cuts in the spectra across a region with strongly increased FWHM at low emission (bottom row) and across a spicule with high emission (top row). The spectra along these cuts (right panels of Fig.~\ref{lim1}) show that the emission pattern monotonically reduces both in amplitude and line width with increasing limb distance. In addition to the single-Gaussian fit, we ran a double-Gaussian fit over these profiles as well. Figure \ref{lim2} shows every second individual profile along the cut of the bottom row of Fig.~\ref{lim1} and the corresponding Gaussian fits. It is obvious that the single-Gaussian fit is not sampling the main emission peak and gets far too broad at about pixel No.~75 (towards smaller numbers). The double-Gaussian fit performs significantly better, but the true emission amplitude for profiles below about No.~69 is difficult to assess because it is comparable to the residual amplitude of, e.g., the photospheric lines.  
\begin{figure}
%\resizebox{12.cm}{!}{\includegraphics{indivspec2.ps}}
\resizebox{12.cm}{!}{\includegraphics{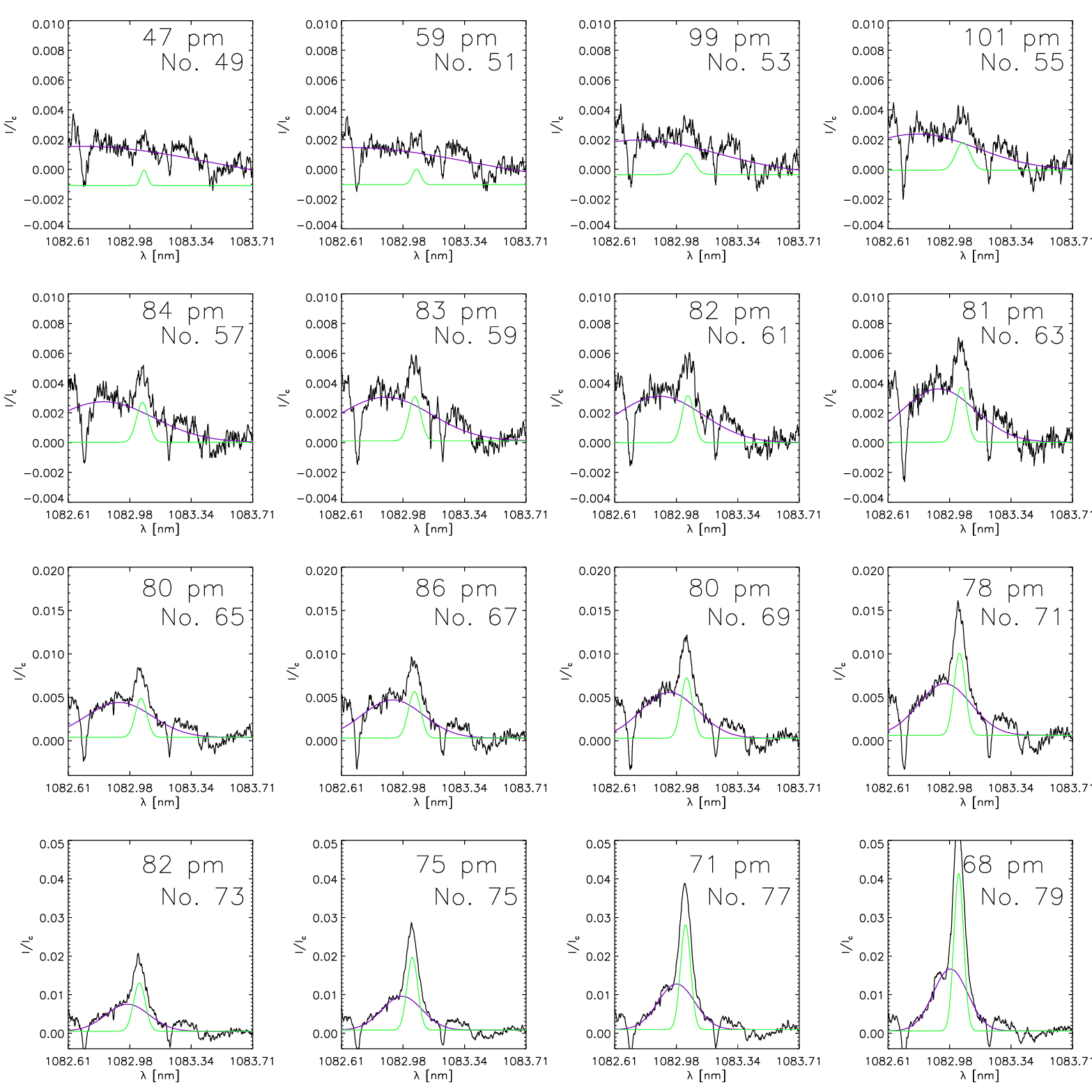}}
\caption{Double-Gaussian fit to individual spectra. The observed spectra along the red bar in the lower row of Fig.~\ref{lim1} are shown with black lines. Green/purple lines show the two components of the double-Gaussian fit. The values at the upper-right corner in each panel give the FWHM of the main (green) component.}\label{lim3}
\end{figure}

The individual components of the double-Gaussian fit are shown in Fig.~\ref{lim3}. One component (green lines) samples the emission peak, while the other captures some broad offset from zero that does not correspond to the blue component of the He emission pattern. For the case of the red emission component, the line width varies around 80\,pm with no clear trend (profiles 57-73), with a slight increase relative to the profiles No.~75-79. The variation comes, however, with the caveat of above: because of the limitations of the data it is not obvious which part of the emission is genuine or spurious.
\begin{figure}
%\resizebox{12cm}{!}{\includegraphics{indivspec4.ps}}
\resizebox{12.cm}{!}{\includegraphics{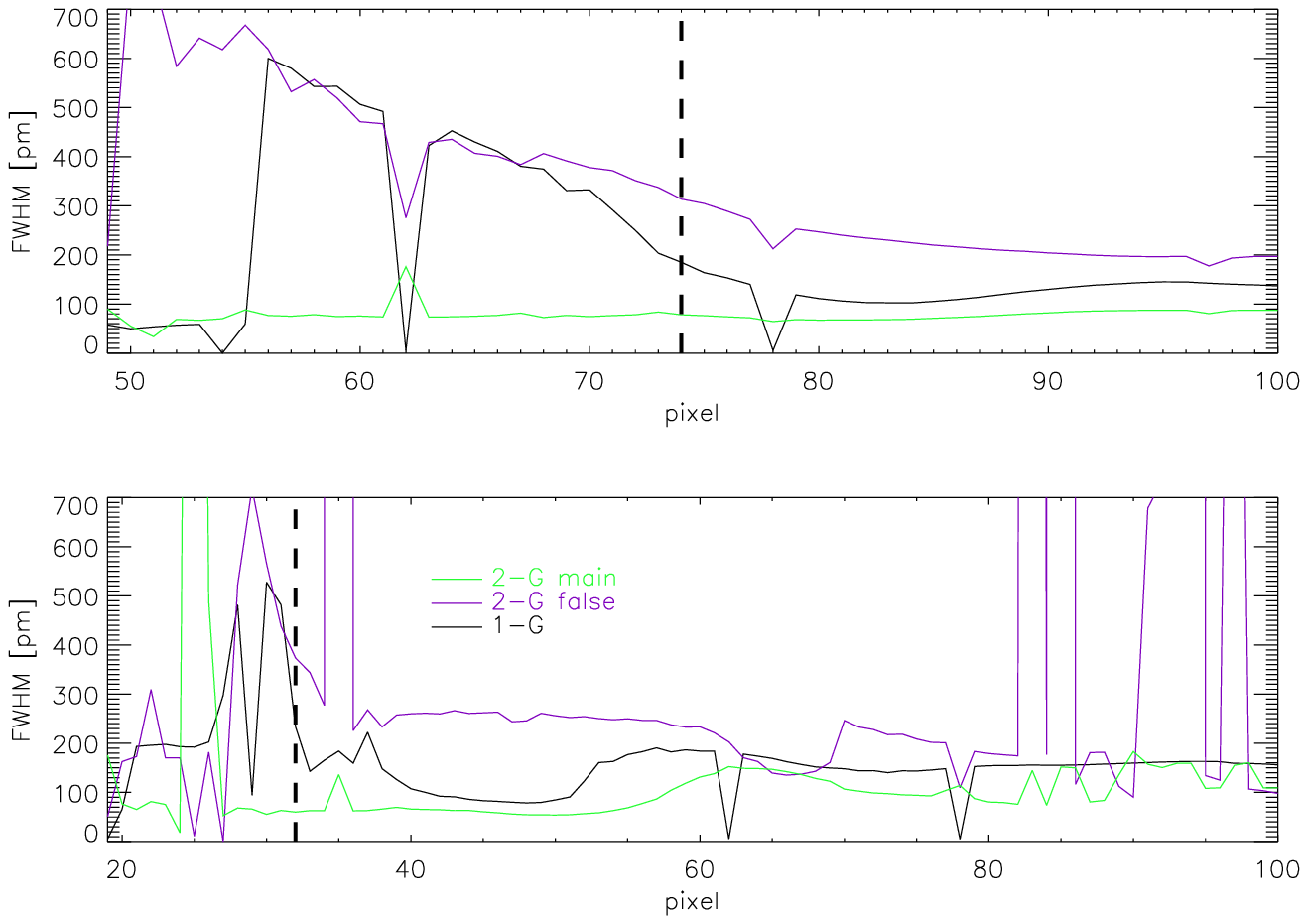}}
\caption{Comparison of FWHM from single and double-Gaussian fits. Top/bottom panel: FWHM of the spectra shown in the bottom/top row of Fig.~\ref{lim1}. Black lines: single-Gaussian fit. Green line: main component of double-Gaussian fit. Purple  line: spurious component of double-Gaussian fit (see Fig.~\ref{lim3}). The vertical black dashed lines denote the maximum extent where the fits are deemed significant and correspond to the blue lines in Fig.~\ref{lim1}. }\label{lim4}
\end{figure}

Figure \ref{lim4} shows the FWHM for both the single and double-Gaussian fits applied to the spectra selected in Fig.~\ref{lim1}. The substantial increase in line width in the single-Gaussian fit can be seen to be fully spurious for the first cut. It is represented by the second, broad component of the double-Gaussian fit that fits the continuum offset. Otherwise the FWHM for the main emission component in the double-Gaussian fit stays roughly constant with a slight reduction in the case of the cut along the spicule. 

Given the limitations of the data and the analysis as discussed above, we thus find the strong increase in FWHM in the He 1083\,nm spectra -- and likewise also the other lines because the line shapes are similar at some height -- to be spurious. In case of low-amplitude profiles, the Gaussian fits might be more misleading than a direct look at the corresponding profiles (Figs.~\ref{spec_temp_gfpi}, \ref{ha_av}, \ref{indiv_spec}, \ref{spic_allwl}, \ref{indi_allwl}, \ref{av_prof_1}, \ref{av_2} and \ref{lim1}). The latter reveals a decrease in line width up to the point where residuals of photospheric blends are as strong as the chromospheric emission.
\subsection{Influence of the blue component at 1082.91\,nm}
\begin{figure}
%\resizebox{12cm}{!}{\includegraphics{indivspec_newtest.ps}}
\resizebox{12cm}{!}{\includegraphics{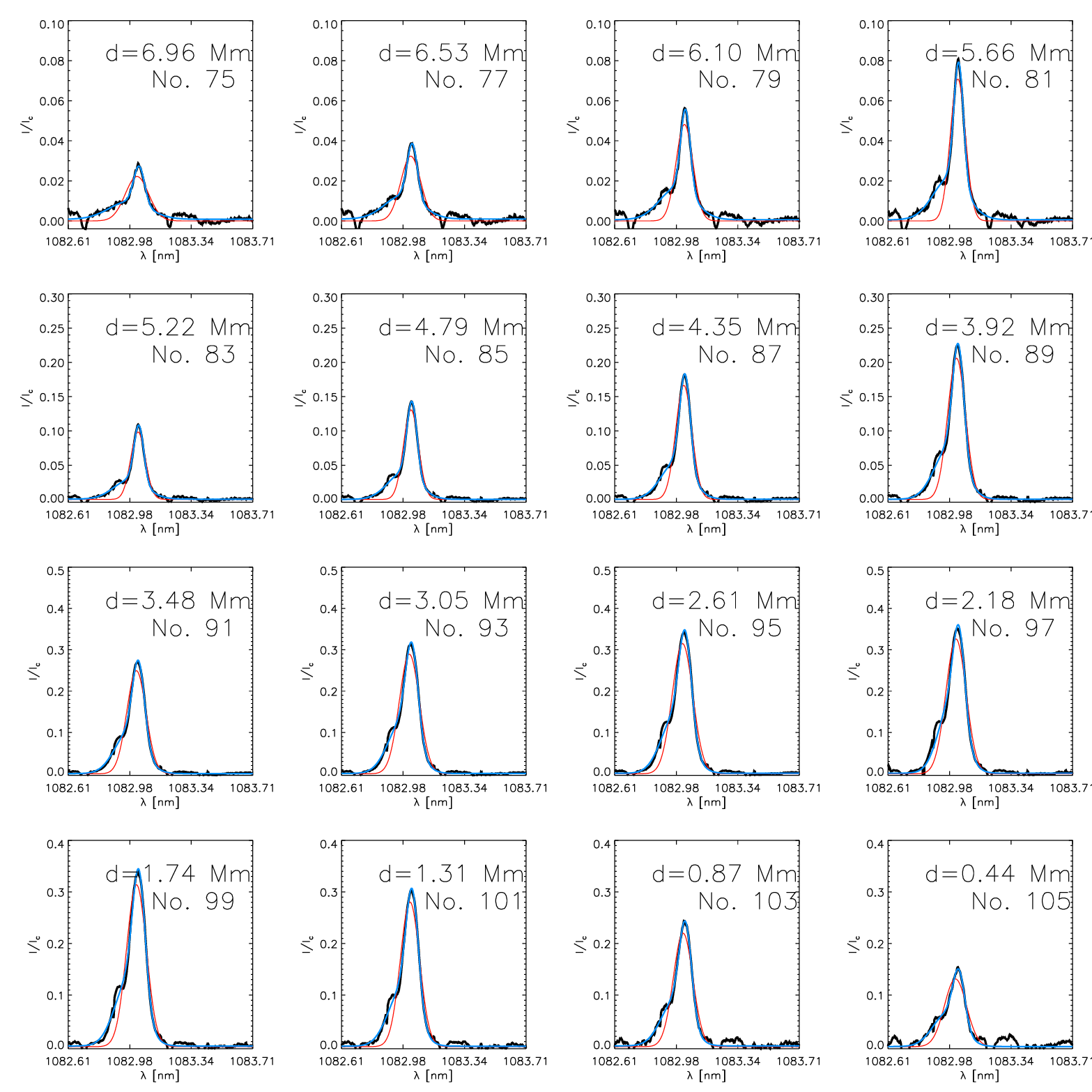}}
\caption{Gaussian fits to individual He I 1083 nm spectra from the limb to a height of about 7\,Mm. The observed spectra are shown with black lines. Red/blue lines show the results of the single/double-Gaussian fits. The limb distance $d$ is given at the top in each panel.}\label{lim6}
\end{figure}
The He line at 1083\,nm consists of two main components at nearby wavelengths at 1082.91\,nm and 1083.03\,nm \citep[see, for instance,][]{sanchezandrade+etal2007}. The ratio of the two components $I_{blue}/I_{red}$ is about 8 in the optical thin case, but can decrease to 3 at a height of about 2\,Mm above the limb  before approaching the optically thin case again at larger limb distances \citep[][their Fig.~4]{sanchezandrade+etal2007}. Figure \ref{lim6} shows that the influence of the weaker blue component on the fit of the red component with a single Gaussian is minor, opposite to the residuals from the stray-light correction discussed in the previous section for profiles with a low light level far from the limb. The line width of the stronger red component of the spectra is well recovered by the single-Gaussian fit for all profiles closer to the limb than $d < 6$\,Mm (profiles No.~80 or more in Fig.~\ref{lim6}; emission amplitude 5\,\% or more). The weaker blue component has little effect on the fit because of its much smaller amplitude. The different amplitude of the components justifies the use of only a single Gaussian for the automated analysis of the spectra across the FOV even if the He line consists of multiple components.
\section{Observations Nos.~2-4 \label{slit_add}}
\begin{figure}
%\centering
%\hspace*{1.5cm}\begin{minipage}{5.cm}
%\includegraphics{FOV_010710.ps}
%\end{minipage}
%\begin{minipage}{2.cm}
%\vspace*{.3cm}\resizebox{1.8cm}{!}{\includegraphics{FOV_010710_zbar.ps}}
%\end{minipage}$ $\\$ $\\$ $\\
%\includegraphics{cuts_010710.ps}
\centerline{\resizebox{7.5cm}{!}{\includegraphics{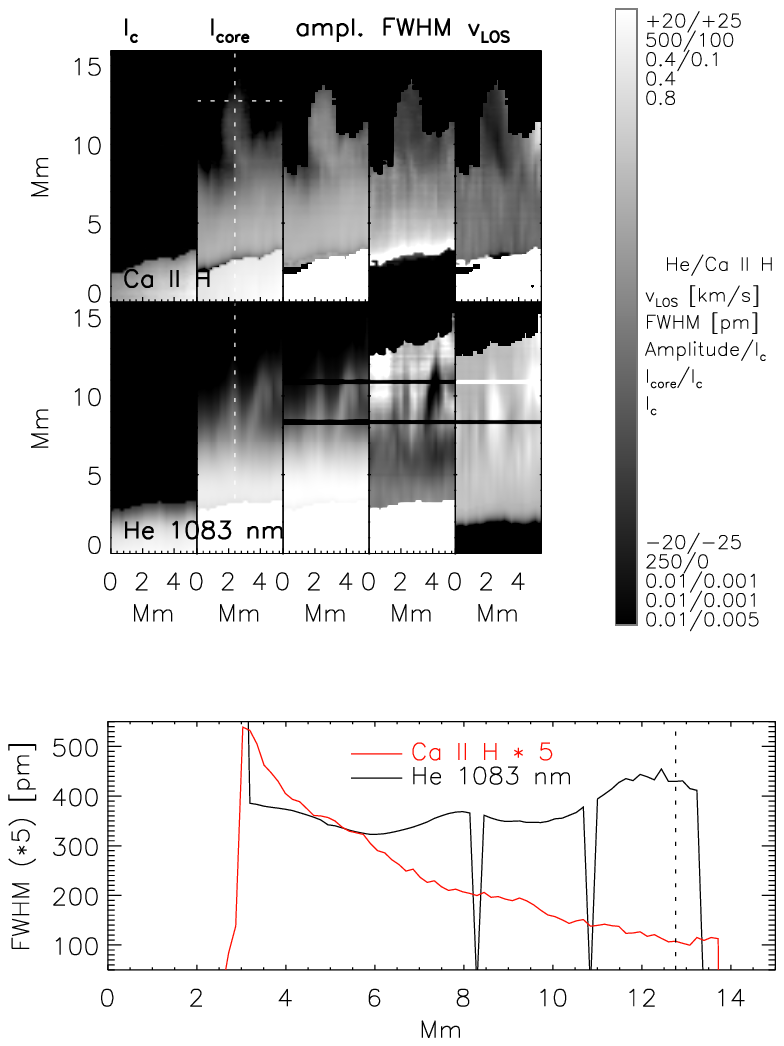}}}
\caption{{Top panel}: results of the Gaussian fit in \ion{Ca}{ii} H and \ion{He}{i} at 1083\,nm for observation No.~2 with a 60-s integration time. {Bottom panel}: FWHM in \ion{Ca}{ii} H and \ion{He}{i} at 1083\,nm along the vertical dotted line in the top panel. The horizontal dotted line in the top panel and the vertical dotted line the bottom panel denote the maximal range with spectra above the noise level.}\label{60secfick}
\end{figure}

The top panel of Fig.~\ref{60secfick} shows the results of the Gaussian fit to the \ion{Ca}{ii} H and \ion{He}{i} spectra of observation No.~2. This observation was selected because of its long integration time of 60\,s that allows one to trace also features of very low intensities. The corresponding H$\alpha$ line-core images in Figs.~\ref{ha_1} and \ref{ha_2} show that during this scan a large-scale (width and height of $\sim 4\,$Mm$\,\times\,$13\,Mm ) structure with a complex temporal evolution moved across the FOV. The slit-spectrograph data captured in this case nothing of the temporal evolution, but only the sheer height extent of the structure up to 13\,Mm above the limb. Thanks to the long integration time, even the spectra far away from the limb could still be reliably analyzed. The plot of FWHM in the bottom panel of Fig.~\ref{60secfick} confirms the visual impression of the top panel that the FWHM in \ion{Ca}{ii} H is monotonically decreasing up to the top of the feature. Above about $y\sim$8\,Mm, the feature presumably is the only one existing at that height, so the monotonic decline of the FWHM should not indicate a possible reduction of the overlap of different features along the LOS. For \ion{He}{i} at 1083\,nm, the FWHM shows a rise at the top, but in the corresponding maps of \ion{He}{i} line-core intensity or the amplitude of the Gaussian, the intensity above 11\,Mm is virtually zero.

\begin{figure*}
%\hspace*{.5cm}\resizebox{12cm}{!}{\includegraphics{FOV_010710_1.ps}}\\$ $\\\centerline{\resizebox{8cm}{!}{\includegraphics{cuts_010710_1.ps}}}
\centerline{\resizebox{11.5cm}{!}{\includegraphics{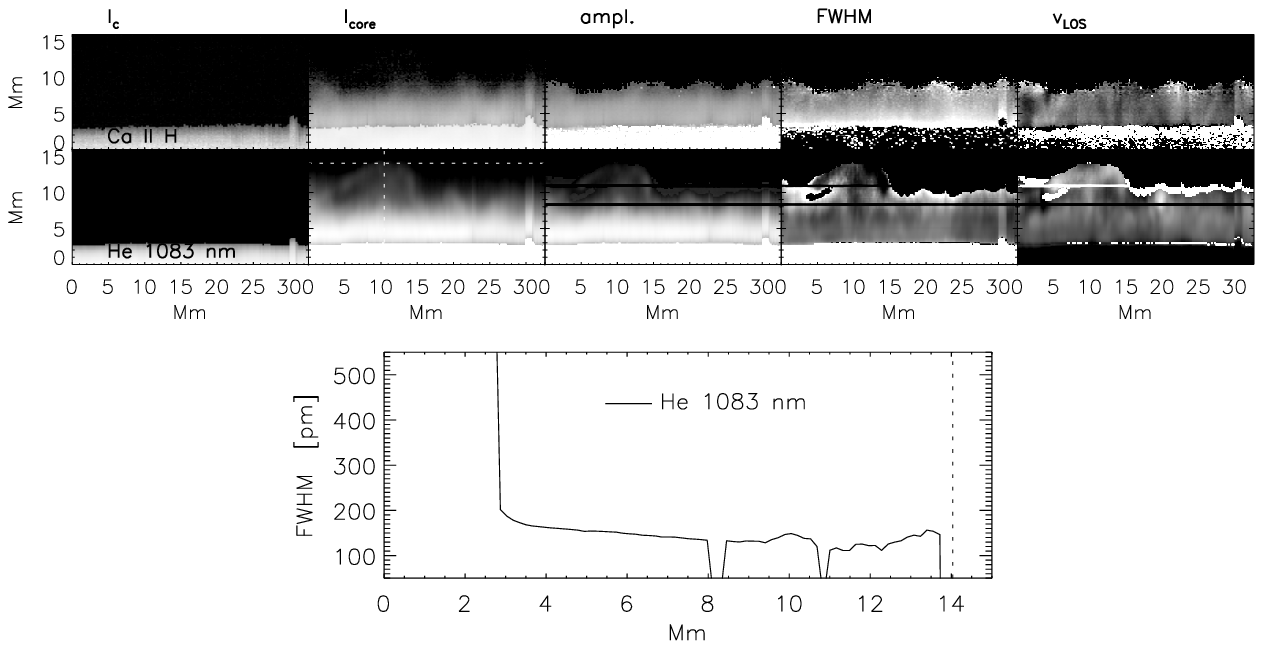}}}
\caption{{Top}: results of the Gaussian fit in \ion{Ca}{ii} H and \ion{He}{i} at 1083\,nm for observation No.~3. {Bottom}: FWHM in \ion{He}{i} at 1083\,nm along the vertical dotted line in the top panel. The horizontal dotted line in the top panel and the vertical dotted line the bottom panel denote the maximal range with spectra above the noise level. For the display ranges see the grey bar of Fig.~\ref{60secfick}.}\label{last_slit_perp}
\end{figure*}

The top panel of Fig.~\ref{last_slit_perp} shows the results of the Gaussian
fit to the \ion{Ca}{ii} H and \ion{He}{i} spectra of observation No.~3, while
the corresponding H$\alpha$ line-core images are shown in Figs.~\ref{ha_3} and
\ref{ha_4}. Similar to observation No.~3, the H$\alpha$ line-core images
exhibit mainly complex-shaped structures without much resemblance to spicules
at heights larger than 5\,Mm above the limb. None of these features are
captured in the \ion{Ca}{ii} H spectra that basically drop to zero intensity
everywhere throughout the FOV above that height (top row of
Fig.~\ref{last_slit_perp}), whereas the \ion{He}{i} spectra still sampled them
up to about 10\,Mm above the limb. The FWHM of \ion{Ca}{ii} H shows a rather
uniform decrease with height above the limb. The FWHM of the \ion{He}{i} line
at 1083\,nm shows again a basically lateral structuring with little vertical
variation. On the location used for a vertical cut in the \ion{He}{i} spectra, the FWHM decreases monotonically up to about 10\,Mm above the limb (bottom panel of Fig.~\ref{last_slit_perp}). 

\begin{figure*}
%\hspace*{.75cm}\resizebox{11.cm}{!}{\includegraphics{FOV_020710_1_mod_new.ps}}$ $\\$ $\\
\centerline{\resizebox{11.5cm}{!}{\includegraphics{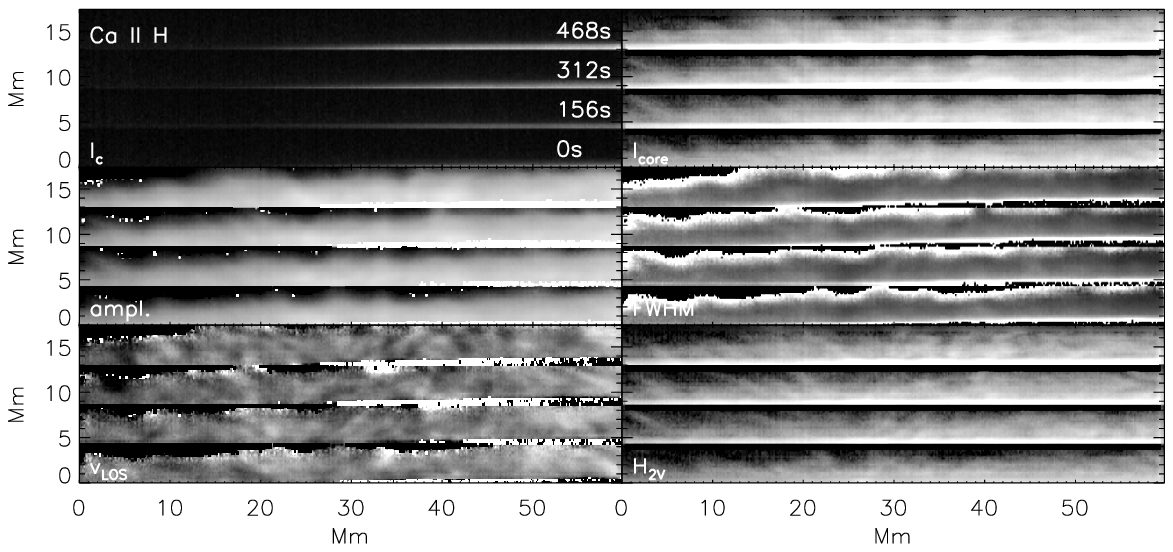}}}
\caption{Results of the Gaussian fit in \ion{Ca}{ii} H for observation No.~4. {Left column, top to bottom}: continuum intensity, amplitude and LOS velocity. {Right column, top to bottom}: line-core intensity, FWHM and intensity of the H$_{\rm 2V}$ emission peak. Note that here $x$ corresponds to the direction along the slit and $y$ to the scanning direction, opposite to the previous figures. The four panels in $y$ for each quantity correspond to four repetitions of the scan taken with a cadence of 156\,s. Display ranges similar as in Fig.~\ref{60secfick}.}\label{slit_parallel}
\end{figure*}

The last example of the 2010 data (observation No.~4,
Fig.~\ref{slit_parallel}) was taken with the slit oriented parallel to the
limb that is located just at the bottom border of the FOV. We could not apply any correction for stray light to these data because of the limited extent of the FOV that neither contains a sufficiently large on-disc region nor a suitable off-limb area for calculating an average profile to be used in the stray-light correction. A few features can be called distinguishable (at $x\sim 20, 28$ and 33\,Mm) in the line-core image or the intensity of the H$_{\rm 2V}$ emission peak. The temporal evolution of their FWHM shows no indication for a significant change of the FWHM with time or height above the limb  within the limitations of cadence and spatial resolution.

\end{appendix}
\end{article} 
\end{document}